\documentclass{gGAF2e}

\begin{document}
	
	
	
	
	
	\title{Experiments and long-term High Performance Computations on amplitude modulations of strato-rotational flows}
	
	\author{{G. Meletti$^{a, b \ast}$\thanks{$^\ast$Corresponding author. Email: gabriel.meletti@b-tu.de}, S. Abide$^c$, S. Viazzo${^b}$, A. Krebs ${^a}$, and U. Harlander${^a}$ \vspace{6pt}}
	\\\vspace{6pt} ${^a}$ Department of Aerodynamics and Fluid Mechanics, Brandenburg University of Technology (BTU) Cottbus-Senftenberg, Cottbus, Germany; ${^b}$Laboratoire de M\'ecanique, Mod\'elisation et Proc\'ed\'es Propre, Aix-Marseille University, CNRS, Centrale Marseille, France; ${^c}$ Laboratoire de Math\'ematiques et Physique, University of Perpignan Via Domitia, Perpignan, France.}
		
	
	\maketitle
	
	\begin{abstract}
    		
    The present paper describes a combined experimental and high performance computing study of new specific behaviors of the Strato-Rotational Instability (SRI). The SRI is a purely hydrodynamical instability that consists of a classical Taylor-Couette (TC) system under stable axial density stratification. The density stratification causes a change on the marginal instability transition when compared to classical non-stratified TC systems, making the flow unstable in regions where -- without stratification -- it would be stable. This characteristics makes the SRI a relevant phenomenon in planetary and astrophysical applications, particularly in accretion disk theory. 
    
    In spite of many advances in the understanding of strato-rotational flows,  the confrontation of experimental data with non-linear numerical simulations remains relevant, since involved linear aspects and non-linear interactions of SRI modes still need to be better understood. These comparisons also reveal new non-linear phenomena and patterns not yet observed in the SRI, that can contribute for our understanding of geophysical flows. 
    		
    The experiment designed to investigate these SRI related phenomena consists of two cylinders that can rotate independently, with the space between these two vertical cylinders filled with a silicon oil.
    For obtaining a stable density stratification along the cylinder axis, the bottom lid of the setup is cooled, and its top part is heated, with temperature differences varying between $3K<\Delta T<4.5K$, establishing an axial linear gradient, leading to Froude numbers $Fr = \Omega_{in}/N$ between $1.5< Fr < 4.5$, where $\Omega_{in}$ is the inner cylinder rotation, and $N$ is the buoyancy frequency.
    The flow field resulting from the cylinders rotation interacting with the stable density stratification is measured using low frequency Particle Image Velocimetry (PIV).  In the present investigations, we focus on cases of moderate Reynolds numbers ($Re$, based on the inner cylinder radius and angular velocities), varying between $Re=300$ and $Re=1300$, and rotation ratio between outer and inner cylinders fixed at $\mu= \Omega_{out}/\Omega_{in} = 0.35$, a value slightly smaller than the Keplerian velocity profile, but beyond the Rayleigh limit.
    The same experimental configuration is also investigated by performing several Direct Numerical Simulations using a parallel high-order compact schemes incompressible code, that solves the Boussinesq equations combining a 2d-pencil decomposition and the reduced Parallel Diagonal Dominant spectral-like method for an efficient parallelization. 
    Both simulations and experiments reveal, in agreement with recent linear stability analyses, the occurrence of a return to stable flows with respect to the SRI when the Reynolds numbers increase.
    Low frequency velocity amplitude modulations related to two competing spiral wave modes, not yet reported, are observed both numerically and experimentally.
	\end{abstract}
	

	\section{Introduction}\label{section:Intorducion}
	
	Understanding the hydrodynamical mechanisms that can result in an outward transport of angular momentum is a central problem regarding stars and planets formation, particularly in the theory of accretion discs \citep{fromang2017angular}. 
	Accretion disks are astrophysical disk-like shape objects composed of gas and dust that rotate around a central object, as a star or a planet. One example of such astrophysical objects is the one observed by the Atacama Large Millimeter/submillimeter Array (ALMA) collaboration \citep{brogan20152014},
	that is an ideal system for the study of disk instabilities and early planet formation, since it consists of a young star surrounded by a disk with high mass. The disk mass $M_d$ is estimated between $0.03M_\odot < M_d< 0.14 M_\odot\ $, 
	and its outer radius is estimated to be $r_{out}\approx 130AU$, with Keplerian velocity profile. At a radius of $r \approx 25 AU$, $\left| u \right|\approx7.0km/s$. The mass of the HL Tauri star found in the center of the disk is estimated to be 30\% higher than the solar mass ($M_* \approx 1.3 M_\odot\ $), enclosed in a radius $r\leq25AU$.
	Central objects in accretion disks, as the HL Tauri star, are formed by the gravitational collapse of the disk matter, but the large mass and sizes values in these systems show that even the slight rotations lead to too much angular momentum \citep{fromang2017angular}, large enough to overcome gravitational forces that would allow the formation of central massive objects. Since astrophysical observations show these massive bodies in the center of accretion disks, the gas flow surrounding the objects should be turbulent, as turbulence, unlike viscous diffusion, can efficiently transport these high angular momentum away from the center of the disk, removing energy from the disk during this process, and allowing gravity to be stronger than the outer-radial angular momentum component, collapsing matter to form the observed astrophysical bodies. 
	
	The gas-dust region of the accretion disks look like a simple differentially rotating shear flow -- known as Taylor-Couette (TC) flows -- with near-Keplerian velocity profiles \citep{Dubrulle2004}.
	A classic TC system consists of two concentric cylinders that rotate with angular velocities $\Omega_{in}$ and $\Omega_{out}$, and has a mean azimuthal velocity profile $\overline{u_\phi}$ given by: 
	\begin{equation}\label{eq:TC}
	\overline{u_\phi(r)} = \Omega_{in}r(\mu-\eta^2) + \frac{r_{in}^2\Omega_{in}(1-\mu)}{r (1-\eta^2)} ,
	\end{equation}
	where $\mu= \Omega_{out}/\Omega_{in}$ is the rotation ratio between inner and outer cylinders, and $\eta=r_{in}/r_{out}$ is the aspect ratio between inner and outer cylinder radius ($r$).
	Equation (\ref{eq:TC}) is the analytical solution of the Navier-Stokes equations in cylindrical coordinates~($\phi$,~$r$,~$z$) for incompressible Newtonian fluids in infinite long cylinders.
	When the first term of right hand side of (\ref{eq:TC}) is zero, the velocity is a potential field, therefore curl free. This defines the Rayleigh limit, $\mu=\eta^2$, that separates stable from unstable flows. 
	
	When $\mu<\eta^2$, $u_\phi(r)$ is unstable and Taylor vortices can be observed. If $\mu>\eta^2$, $u_\phi(r)$ remains stable.
	In accretion disks, the Keplerian azimuthal angular velocity profile $\Omega(r)\propto r^{-3/2}$ \citep{Dubrulle2004} leads to $\mu=\eta^{3/2}$, and hence to stable velocity profiles ($\mu>\eta^2$). This raises the question of which mechanisms could destabilize these rotating gas flows generating the turbulent outward angular momentum transport. 
	
	Among other candidates, the Strato-rotational Instability (SRI) has attracted attention in recent years as a possible instability leading to turbulent motion in accretion disks \citep{Dubrulle2004, lyra2019initial}. In contrast to the Magnetorotational Instability (MRI), the SRI is a purely hydrodynamic instability consisting of a classical Taylor-Couette (TC) system with stable density stratification due to, for example, salinity \citep{Withjack, Boubnovt1995,ParkJunho2013,LeBars2007}, or to a vertical temperature gradient \citep{rudiger2017stratorotational, seelig2018experimental}.
	\cite{Dubrulle2004} conclude that, in  astrophysical disks, stable stratification due to temperature differences is the rule rather than the exception. 
	When a stable stratification is imposed to the TC system, the flow can be destabilized for $\mu>\eta^2$ leading to the SRI. 
	
\begin{center}
		\begin{figure}[!t] 
			\begin{minipage}[t]{0.32\linewidth}
				\centering
				\includegraphics[trim={0cm 0cm 0cm 0cm},clip, width=1 \linewidth]{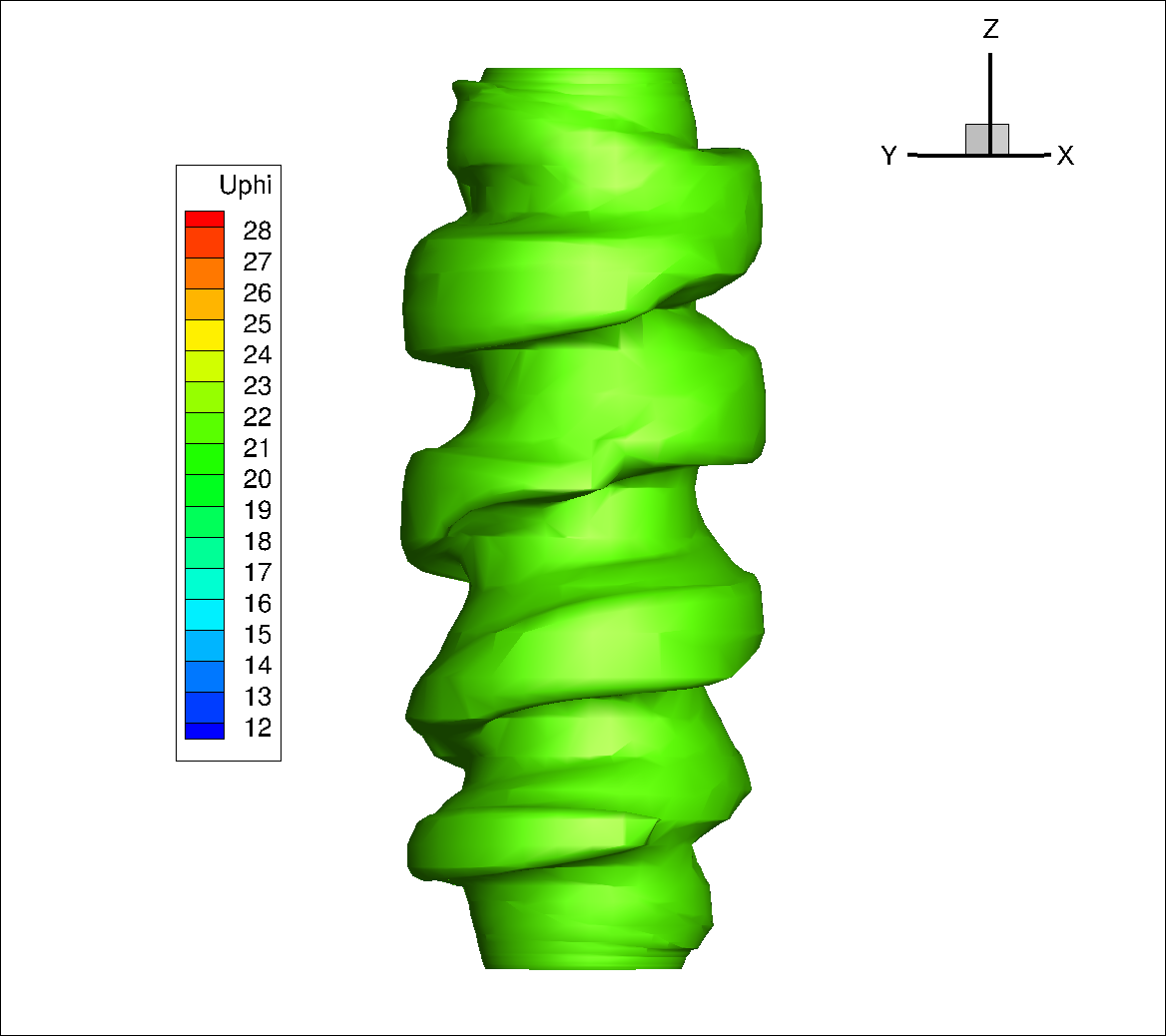}
				\caption*{(a) ${u}_\phi$} 
				\vspace{4ex}
			\end{minipage}
			\hspace{.01in}
			\begin{minipage}[t]{0.32\linewidth}
			\includegraphics[trim={0cm 0cm 0cm 0cm},clip, width=1 \linewidth]{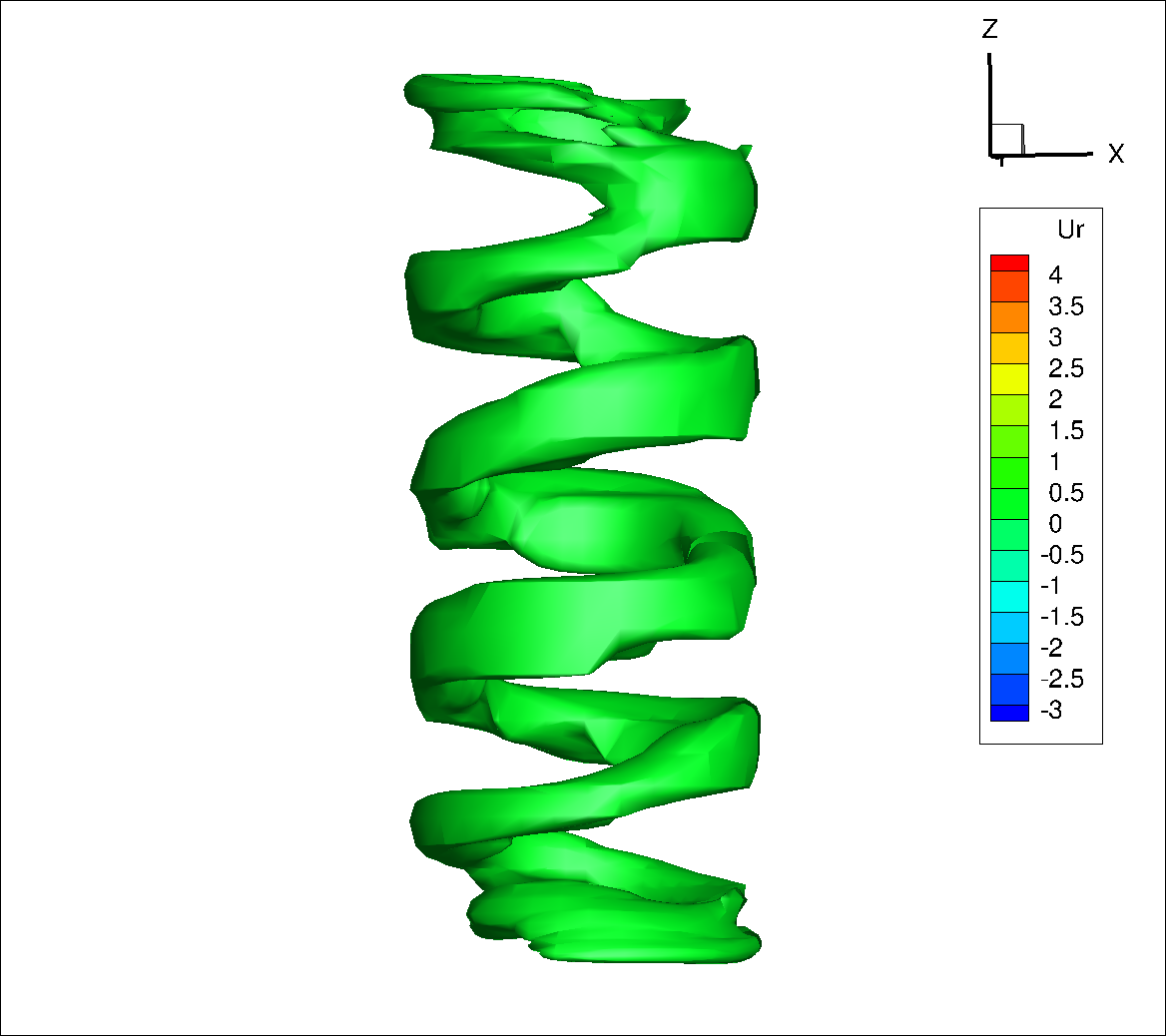}
				\centering
			\caption*{(b) ${u}_r$} 
			\vspace{4ex}
		\end{minipage} 
		\hspace{.01in}
		\begin{minipage}[t]{0.32\linewidth}
			\centering
			\includegraphics[trim={0cm 0cm 0cm 0cm},clip, width=1 \linewidth]{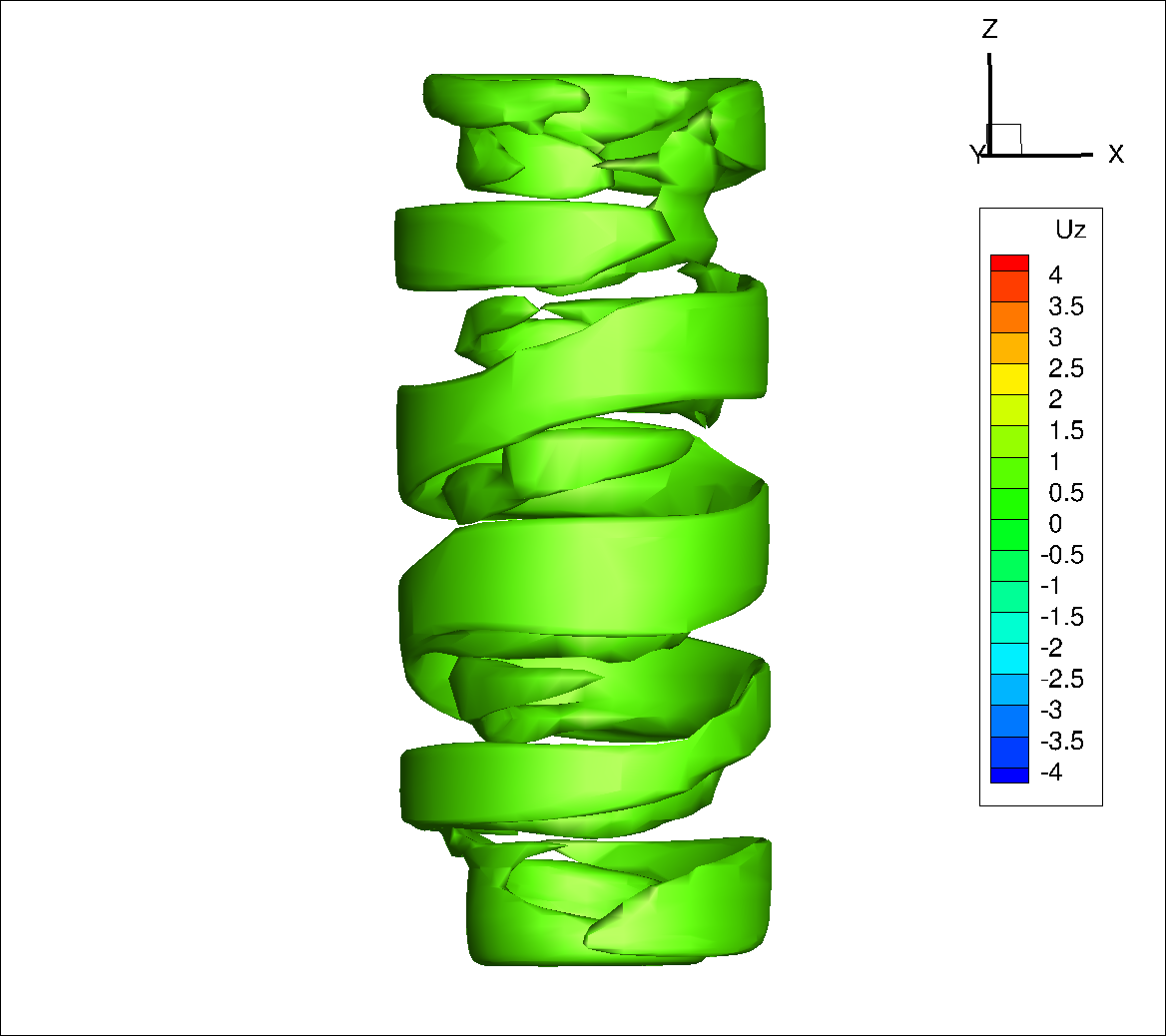}
			\caption*{(c) ${u}_z$} 
			\vspace{4ex}
		\end{minipage} 
		\caption{Velocity isosurfaces snapshots of the SRI showing its non-axisymmetric spirals obtained by numerical simulation with $\mu=0.35$, $\eta=0.517$, a linear stable axial temperature gradient with $\partial T/\partial z \approx 5.7K/m$, $Fr\approx1.5$ and $Re=400$. The aspect ratio between cavity height and gap is \mbox{$\Gamma = H/(r_{out}-r_{in})=10$}. No-slip and impermeable Dirichlet boundary conditions are imposed ($u_\phi(r_{in})=\Omega_{in}r_{in}$, $u_\phi(r_{out})=\Omega_{out}r_{out}$, $u_z(z=0)=0$, $u_z(z=H)=0$, $u_r(r_{in})=0$, $u_r(r_{out})=0$), and the bottom and top lids of the cavity rotate with the outer cylinder rotation $\Omega_{out}$.} \label{fig:tecplots}
	\end{figure}
\end{center}

	Experiments performed by \cite{Withjack} found that, unlike the axisymmetric rolls in classic TC flows, the SRI presents non-axisymmetric spirals, confirmed by later experiments performed by \cite{Boubnovt1995}, as an example. Figure \ref{fig:tecplots} shows isocontours of azimuthal ($u_\phi$), radial ($u_r$), and axial ($u_z$) velocity components where these non-axisymmetric SRI spiral structures can be observed. Note that the spirals in figure \ref{fig:tecplots} were obtained in a region where the flow is stable with respect to the non-stratified TC regime.


\cite{caton2000stability} showed by linear stability analysis and experimental observations that the first SRI transition happens via a supercritical Hopf bifurcation that destabilizes purely azimuthal flows. \cite{rudiger2017stratorotational} obtained marginal stability curves using linear stability analysis for different values of $Rn$, the Reynolds number based on the buoyancy frequency, defined as:
\begin{equation}\label{eq:Rn}
Rn = N r_{in} (r_{out}-r_{in})/\nu .
\end{equation}
where $\nu$ is the kinematic viscosity of the fluid, and $N$ is the buoyancy frequency, also known as Brunt-V\"ais\"al\"a frequency
\begin{equation}\label{eq:Brunt-Waisailla}
N = \sqrt{\alpha g \frac{\partial T}{\partial z}} .
\end{equation}
Here, $\alpha$ is the coefficient of thermal expansion, g is the gravity constant, and $\partial T/\partial z$ is the axial temperature gradient. Note that, higher temperature gradients lead to higher values of $N$, and consequently, of $Rn$.

The Reynolds number of the stratified Taylor-Couette flows, based on the inner cylinder rotation ($\Omega_{in}$), is defined as:
\begin{equation}\label{eq:Reynolds_number}
Re = \Omega_{in} r_{in} (r_{out}-r_{in})/\nu .
\end{equation}

The Froude number measures the relative importance of rotation and stratification, being defined as:
\begin{equation}\label{eq:Fr}
Fr = \frac{Re}{Rn} = \frac{\Omega_{in}}{N} .
\end{equation}

\begin{figure}[tp!]
	\centering
	\includegraphics[width=.6\linewidth]{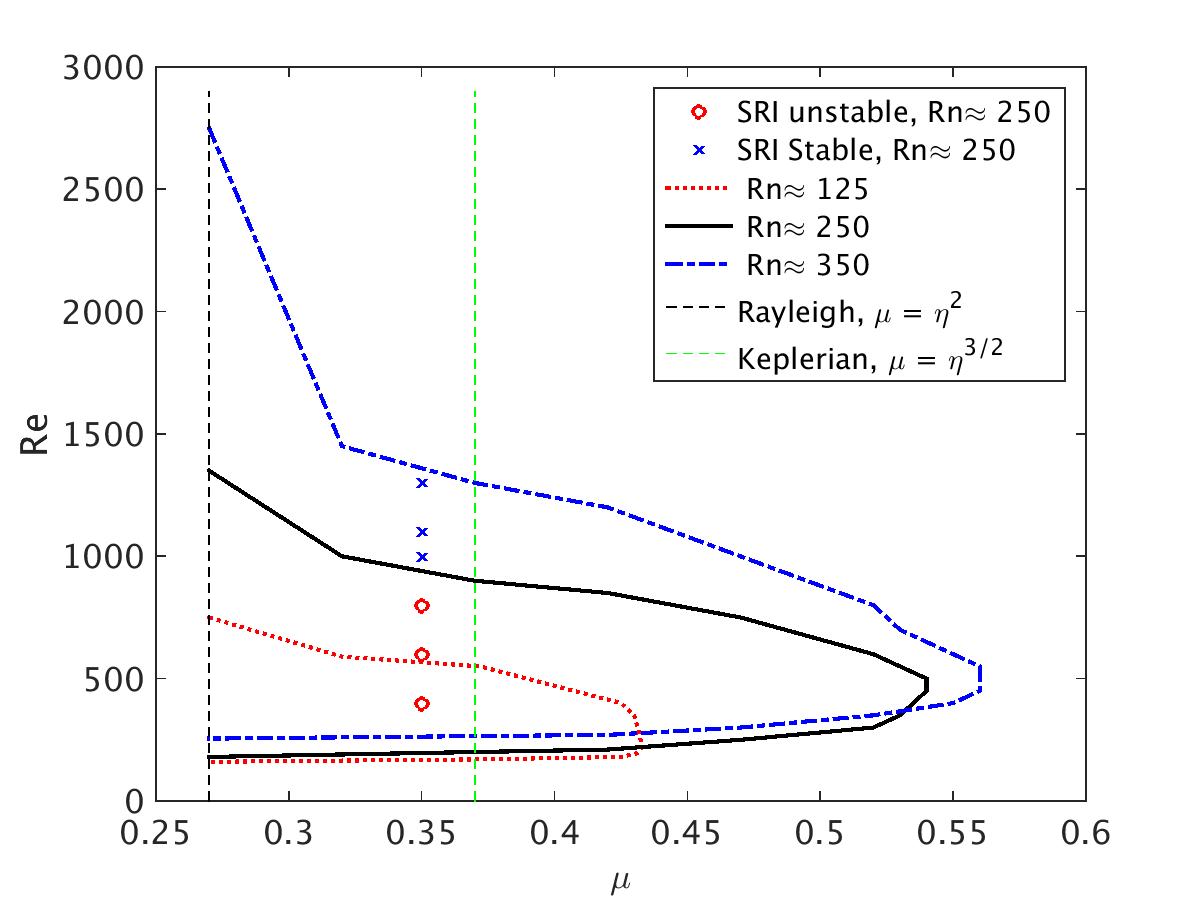}
	\caption{SRI marginal linear stability curves redrawn from \cite{rudiger2017stratorotational} for 3 different values of $Rn$. Bigger values of $Rn$ (and $N$) lead to larger instability regions. The black dashed vertical line on the left represents the Rayleigh limit $\mu=\eta^2$, that separates stable and unstable non-stratified TC cases (any to the left of this line in the diagram would be TC unstable). The green dashed vertical line on the right represents the Keplerian line $\mu=\eta^{3/2}$. SRI unstable (red circles) and stable (blue crosses) cases with $Rn\approx250$, $\mu=0.35$ with different Reynolds numbers were investigated numerically and experimentally in this article.} \label{fig:linear_stability_curve}
\end{figure}

Figure \ref{fig:linear_stability_curve} shows marginal stability curves redrawn from \cite{rudiger2017stratorotational} for different $Rn$ values, and for the configurations we studied numerically and experimentally. 
Flows inside (outside) these curves are predicted to be SRI unstable (stable). Note that the SRI unstable regions increase with $Rn$.
Obviously, stratified flows with $\mu>\eta^2$ can be unstable (to the right of the Rayleigh limit, shown by the black vertical dashed line). Most important, the Keplerian profile relevant for accretion disks can be unstable for all chosen $Rn$.
The unstable regions in figure \ref{fig:linear_stability_curve} were confirmed by experiments \citep{rudiger2017stratorotational,seelig2018experimental}, but the upper transition back to stable regimes was not captured clearly in the previous experiments in the region $0.3<\mu<0.4$. On the other hand, \cite{edlund2014nonlinear,edlund2015reynolds} and \cite{lopez2017boundary} studied turbulent boundary layer instabilities in quasi-Keplerien flows, and observed numerically and experimentally that increased Reynolds numbers relaminarize the flow, in agreement with the predictions of linear stability theory \citep{rudiger2017stratorotational}. Following \cite{lyra2019initial}, these uncertainties show the importance  of  understanding the  characteristics of stratified TC flows at increasingly higher Re values, highlighting the dangers of deriving conclusions regarding the behavior of protoplanetary disks based on relatively low $Re$ quasi-Keplerian experiments and numerical simulations.

The SRI spirals are different from the ones observed in non-stratified TC flows. \cite{Yim2015} investigated the reasons why the columnar vortex shape of strato-rotating fluids are different from the shear, centrifugal or radiative instabilities. \cite{leclercq2016connections} investigated connections between TC flows, radiative instabilities and the SRI.
\cite{hoffmann2009nonlinear} showed changes in the Taylor vortices of non-stratified TC systems when a low Reynolds number flow is forced in the axial direction. In these cases, the rolls assume inclination and shape similar to the ones observed in the SRI, traveling along the axial axis in the same direction as the flow imposed. These TC spiral propagation and inclination is affected by non-linear defects that are also observed in the SRI. When the external axial flow is stopped, there is a break of symmetry associated to a Hopf bifurcation, and a pattern change occurs, with changes in the spiral inclination, but keeping the same spiral shape observed in the SRI flow.

In the following section \ref{section:Exp_setup}, the experimental setup is described together with the particle image velocimetry (PIV) system we used for measurements in the azimuthal-radial ($\phi-r$) plane. In section \ref{section:Numerical_simulations_desription}, we present the DNS code, based on a combination of higher-order accuracy and high performance computing.
Comparisons of numerical and experimental data are carried out in section \ref{section:Results}.
In section \ref{section:low_freq_modulations}, we describe low-frequency amplitude modulations and connect those to pattern changes in the SRI spirals. 
Finally, in section \ref{section:conclusions}, we present the conclusions.

%

\section{Experimental Setup}\label{section:Exp_setup}

\begin{figure}[tp] 
	\begin{center}
		\begin{minipage}[tp]{0.45\linewidth}
			\centering
			\includegraphics[trim={3cm 0 2cm 0},clip, width=1\linewidth]{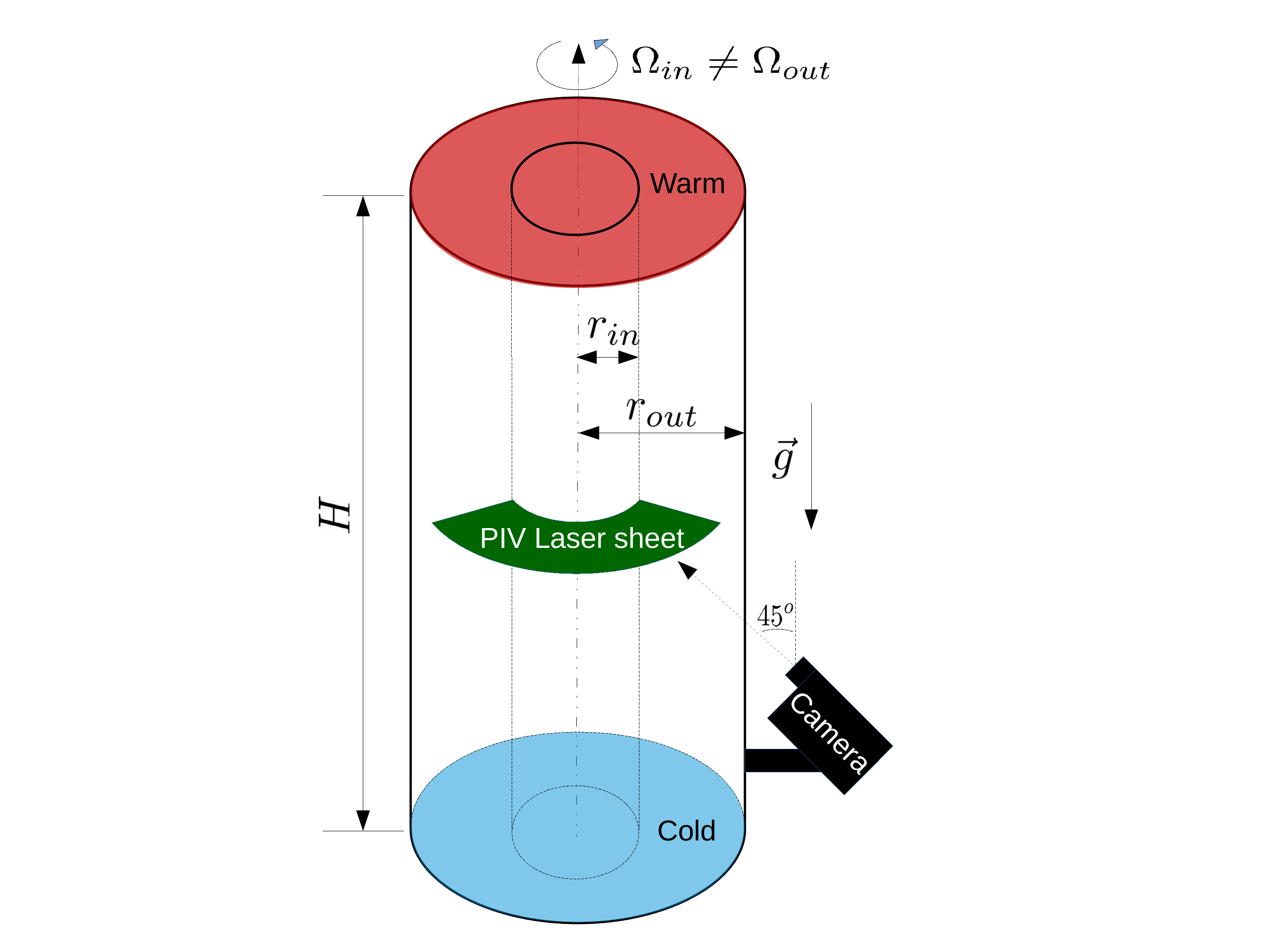}
			\caption{Schematic representation of the SRI experimental setup.} \label{fig:Schematic}
		\end{minipage}
		\begin{minipage}[tp]{0.55\linewidth}
			\centering
			\includegraphics[trim={5cm 9cm 3cm 5cm},clip, angle=0,width=.47\linewidth]{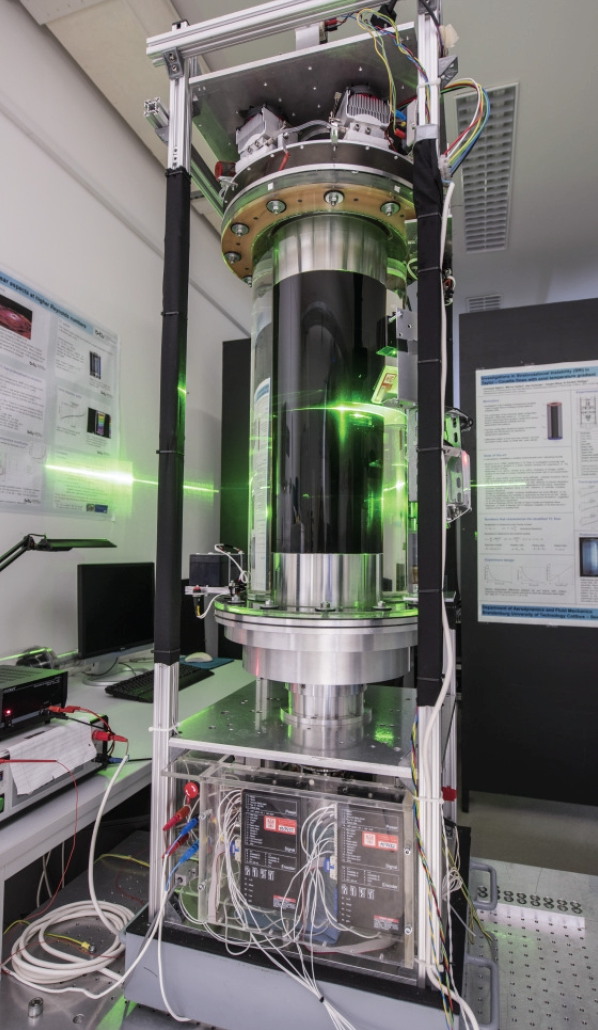}
			\caption{Experimental Setup.}  \label{fig:Pic_ExpSetup}
		\end{minipage}
	\end{center}
\end{figure}

The experimental setup designed for studying SRIs consists of a Taylor-Couette system where the top lid is heated, and the bottom lid is cooled for obtaining a stable density stratification in the axial~($z$)~direction, as schematically represented in figure \ref{fig:Schematic}.

The inner and outer cylinders of the setup are concentric and rotate independently, respectively with angular velocity $\Omega_{in}$ and $\Omega_{out}$, driven by two different motor units controlled by servo amplifiers. The top and bottom parts are closed and connected to the outer cylinder, so both lids rotate with angular velocity $\Omega_{out}$. The gap between inner and outer cylinders is filled with a Newtonian M5 silicon oil of viscosity 5 times higher than the viscosity of water, $\nu(25\degree C)=5.10^{-6}m^2/s$, and has similar specific weight $\rho(25\degree C)=923 kg/m^3$. The higher viscosity of the oil allows us to achieve smaller Reynolds number values compared to water. The coefficient of thermal expansion of the M5 oil at $25\degree C$, used to compute the buoyancy frequency $N$, is $\alpha = 1.08\times10^{-3}/K$. The thermal conductivity, specific heat at constant pressure, and Prandtl number at $25\degree C$ are, respectively, $k=0.133W/K m$, $c_p = 1630 J/kg K$, and $Pr = \frac{\nu}{k/(c_p \rho)} \approx  57$. Hollow glass spherical particles with mean diameter between $10\mu m<d<20\mu m$ and density over water density of~$1.05$ were employed as passive PIV tracing particles. The outer cylinder wall, with radius $r_{out}= 145mm$, is made of transparent acrylic material, to allow optical access to the flow that develops in the cavity between inner and outer cylinders. The inner cylinder radius measures $r_{in} = 75mm$, so the distance between inner and outer cylinders is \mbox{$d=r_{out}-r_{in}=70mm$} . The cylinders height is $H=700mm$, and its geometric parameters are the aspect ratio $\Gamma = H/(r_{out}-r_{in}) = 10$, and the radii ratio $\eta= r_{in}/r_{out} \approx 0.52$.

A mini-\textit{Particle image velocimetry (PIV)} system is used to acquire instantaneous velocity fields in azimuth-radial cross-section. The camera employed in is a \textit{GoPro Hero 4 black edition} with spatial resolution of $1080 \times 1920 px$, acquiring images at a frequency of 24~frames~per~second. A continuous green laser ($\lambda =532 nm$) is employed for performing the measurements, that produces a 2mm thick horizontal light sheet (in the $r-\phi$ plane). All the results presented were performed with the laser at the mid height axial position $H/2=345mm$. The field of view allows the observation of $\Delta \phi \approx 65\degree$ angle in the azimuthal direction, i.e., approximately 18\% of the full horizontal ($\phi-r$) cross section.

The inner cylinder of the experimental setup is made of aluminium and anodized to minimize undesired laser reflections at the inner wall. Even with the dark colour of the ionized aluminium, and because of laser reflection and refraction when the light passes from the acrylic material to the oil at the outer cylinder, the PIV measurements near the walls become spurious, so that, from the full radius of the gap, experimental data are only considered in the interval $80mm<r<143mm$.

The top and bottom end-plates that close the experimental apparatus are also made of aluminium, and cannot be optically accessed during the measurements. The PIV images are then obtained through the outer cylinder wall, with the camera inclined about $45\degree$ with respect to the laser sheet, as represented in figure \ref{fig:Schematic}. To correct the parallax effects generated by the camera inclination, a calibration grid with square chessboard structure is placed at the laser sheet position. The squares of the dashboard appear distorted in the images obtained, and when they are corrected back to their square shape, an undistortion map is created, that is applied to all PIV images. This transformation is performed using a $5^{th}$-order polynomial distortion method in a parallel code described in \cite{seelig2018experimental}. The undistorted chessboard image is then used for calibration, converting the PIV images information from pixels to millimeters. The origin of both concentric inner and outer cylinders is defined in the PIV pictures for transforming the data obtained in Cartesian coordinates to polar coordinates.

The PIV instantaneous velocity fields are computed in the undistorted images using a parallel version of MatPIV \citep{sveen2004introduction}. We used interrogation windows of $32 px \times 32 px$ with 50\% of overlap. This corresponds to a spatial resolution of $1.6 mm$. The vector fields obtained from the PIV measurements are interpolated using moving average with a window size of 25 vectors.

Because the camera field of view does not access the full $360\degree$ of the $\phi-r$ plane, and due to the large amount of data obtained in each instantaneous vector field, the experimental results are stored in time-series matrices for a fixed angle $\phi=0$, at $10^3$ different radial positions.

For heating the experiment's upper lid, twelve TEC263 Peltier elements are installed equidistant from each other at the inner part of the top plate. The advantage of using temperature stratification for experimentally investigating the SRI, instead of salt stratification \citep{Withjack,Boubnovt1995,Shalybkov2005, LeBars2007}, is that the boundary condition can restore the stratification after turbulent mixing effects.
The disadvantage is the time required for establishing a near linear temperature
profile to have a constant buoyancy frequency $N$. The procedures for establishing the temperature profile and for starting the experiment are described in more details in appendix \ref{Appendix:Exp_procedure}.
In the experiments presented in sections \ref{section:Results} and \ref{section:low_freq_modulations}, the temperature differences between top and bottom lids are $3K<\Delta T < 4.5K$.

The rotation ratio between the angular velocity of outer and inner cylinders $\mu=\Omega_{out}/\Omega_{in} $ can be set to different values in our experiments, from counter rotation regimes ($\mu<0$) to co-rotating cases ($\mu>0$). Since the rotation ratio in accretion disks can also be slightly sub-Keplerian \citep{visser2010sub, lyra2019initial}, the results presented are obtained at $\mu \approx 0.35$. This value is smaller than the pseudo-Keplerian line, found at $\mu=\eta^{3/2}\approx0.372$ for our experimental setup, and greater than the Rayleigh line at $\mu = \eta^2 \approx 0.275$. Therefore, at the $\mu$ value chosen, the flow is stable with respect to non-stratified TC (see Fig. \ref{fig:linear_stability_curve}). Note that similar results have been obtained for $\mu\approx0.372$, i.e. at the Keplerian-line, but are not shown here.

\section{Numerical method}\label{section:Numerical_simulations_desription}

To carry out numerical investigations of the SRI, Direct Numerical Simulations were performed using the method reported in \cite{ABIDE2018}, dedicated to high performance computing. In the following, the numerical method is briefly described.\newline
The physical model is the Taylor-Couette flow configuration filled with an incompressible fluid endowed with a vertical temperature gradient. 
Using the Boussinesq approximation to account for the buoyancy forces, the governing equations read:
\begin{equation}\label{eq:Numeric_Eqn}
\left\{
\begin{array}{ll}
\nabla . \textbf{u} = 0 & \text{ in }  D, \\
\partial_t \textbf{u} + \frac{1}{2}\left[(\textbf{u} . \nabla ) \textbf{u} + \nabla . (\textbf{u}\textbf{u}) \right] = - \nabla p + \nu \Delta \textbf{u} +\mathbf{F} & \text{ in }  D,\\
\partial_t T + \frac{1}{2}\left[(\textbf{u} . \nabla ) T+ \nabla . (\textbf{u} T) \right] = \kappa \nabla^2 T & \text{ in }  D,\\
\end{array}
\right.
\end{equation}
where $D$ is the computational domain, $\kappa$ the fluid thermal conductivity, $p$ is the pressure, $T$ is the temperature field, and $\textbf{u}=(u_r,u_\phi,u_z)$ is the velocity vector field in radial, azimuthal and axial directions, respectively.

The body force F is the buoyancy force driven by density variation:
\begin{equation}
F = \alpha \textbf{g} \Delta T .    
\end{equation}
The velocity is prescribed at the walls of the cylinders. 
The vertical temperature gradient results from the temperature difference $\Delta T$ imposed between the top and bottom lids, while considering adiabatic lateral walls.
Note that, recently, \cite{lopez2020impact} included also centrifugal buoyancy forces in the body force term of their SRI numerical investigations, and showed that it had a small contribution in their simulations.%
Although relevant in the equations, the influence of the term is small, and it assumes a value even one order of magnitude smaller in our simulations and experiments. Therefore, its contribution was not included in our simulations. Our model, instead, follows the same approach as \cite{rudiger2017stratorotational}.
The time discretization of (\ref{eq:Numeric_Eqn}) is made using the second order semi-implicit Adams-Bashforth/Backward-Euler scheme. The semi-discretized version leads to a coupled velocity/pressure system for which the diffusive terms are implicit in time. The velocity-pressure coupling is solved using the improved projection method proposed by \cite{hugues1998improved}, thus the discretized Navier-Stokes equations reduce to two Poisson problems for a preliminary pressure and a pressure correction, and three Helmholtz problems arising from the momentum equations.
Higher-accuracy discretization is presently achieved through the spectral Fourier discretization in the azimuthal direction and the fourth-order compact finite difference schemes \citep{abide20052d} in the two other directions. 

The compact schemes have a higher accuracy and a better resolution at high wavenumbers than the centered finite differences for the same stencil size. This property results from the implicit definition inherent to compact schemes. Practically speaking, the compact scheme computation of the time explicit terms is performed with a cost of similar order of magnitude than using centered finite differences. This is no longer true for the implicit  terms, viz the Poisson or Helmholtz equation, since, in this case, the resulting linear system is dense and solvers have to be designed. In the present DNS-code, the direct solver successive diagonalization method is considered.
To benefit from the modern High Performance Computing framework, two parallel strategies have been
implemented. The first one concerns the computation of compact scheme derivative/interpolation which are based on the reduced Partial Diagonal Dominant (rPDD) algorithm of \cite{sun1995application}. This method of solving tridiagonal linear systems shows interesting parallel performances \citep{ABIDE2017} in the context of fluid flow solvers.
The diagonal dominance is exploited to derive an approximate solver involving only neighbor-to-neighbor communications, greatly improving the parallel efficiency. The second strategy concerns the solution of Poisson/Helmholtz equations. Specifically, a 2d-pencil decomposition is considered to get a parallel version of the full diagonalization method \citep{ABIDE2017}. Despite the large communications involved by the global parallel transposes, this approach allows a significant reduction of the simulation time.

The same geometry as presented in section \ref{section:Exp_setup} describing the experimental setup is implemented into the numerical code.

\section{Model validation\label{section:Results}}

In this section, we present a comparison between experimental and numerical simulation data in the radial-azimuthal ($r-\phi$) plane. The objective of this comparison is not only the numerical code validation, since it has already been validated in previous works \citep{ABIDE2017, ABIDE2018}, but also to explore new physical phenomena associated with the SRI that can lead to a better understanding of this still not fully comprehended hydrodynamic instability.
\subsection{Comparison of experimental and numerical SRI data} \label{subsection:Num_exp_comparison}

\begin{figure}[tp] 
	\begin{center}
		\begin{minipage}[t]{0.45\linewidth}
			\centering
			\includegraphics[width=1\linewidth]{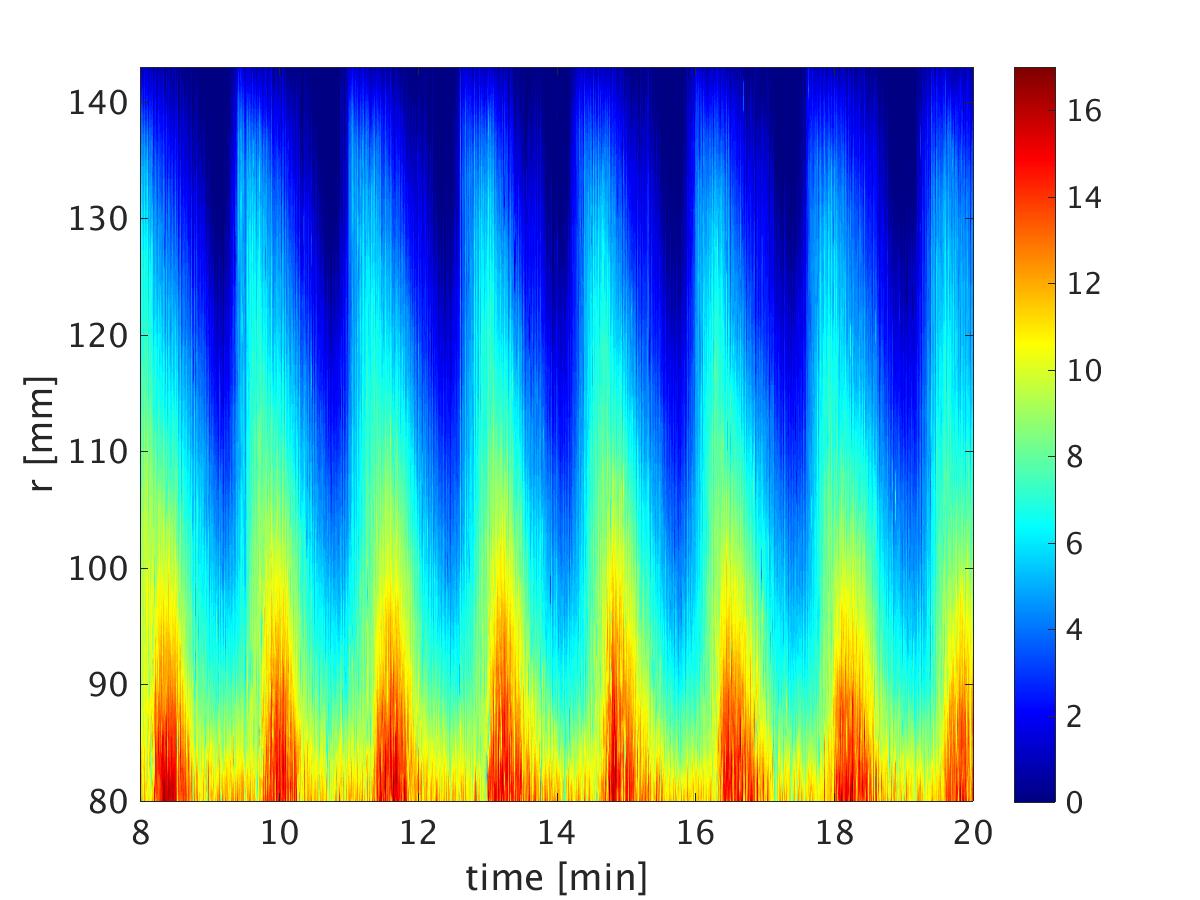}
			\caption*{(a) Experiment} 
			\vspace{4ex}
		\end{minipage}
		\hspace{.1in}
		\begin{minipage}[t]{0.45\linewidth}
			\centering
			\includegraphics[width=1\linewidth]{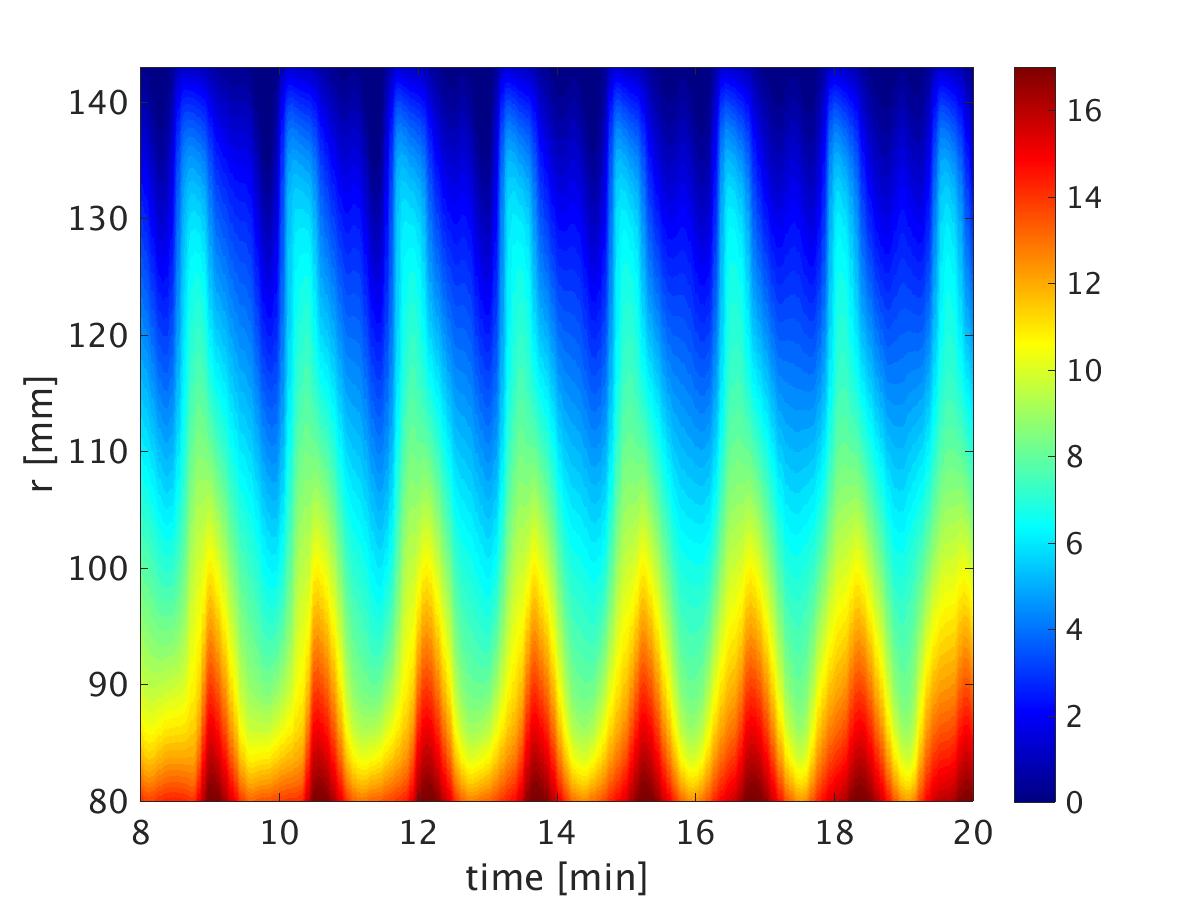}
			\caption*{(b) Numerical Simulation} 
			\vspace{4ex}
		\end{minipage} 
		\caption{$u_\phi$ space time diagram for $Re=400$ in a reference frame co-rotating with the outer cylinder at mid-height axial position ($H/2$). The horizontal axis shows time in minutes and the vertical axis, the radius in mm (the bottom of the image is the inner cylinder region, and the top, the outer cylinder). Both figures (a) and (b) show 12 minutes of measurements.} \label{fig:Uphi_Hovmoller_comparison}
	\end{center} 
\end{figure}

\begin{figure}[tp] 
	\centering
	\includegraphics[width=.5\linewidth]{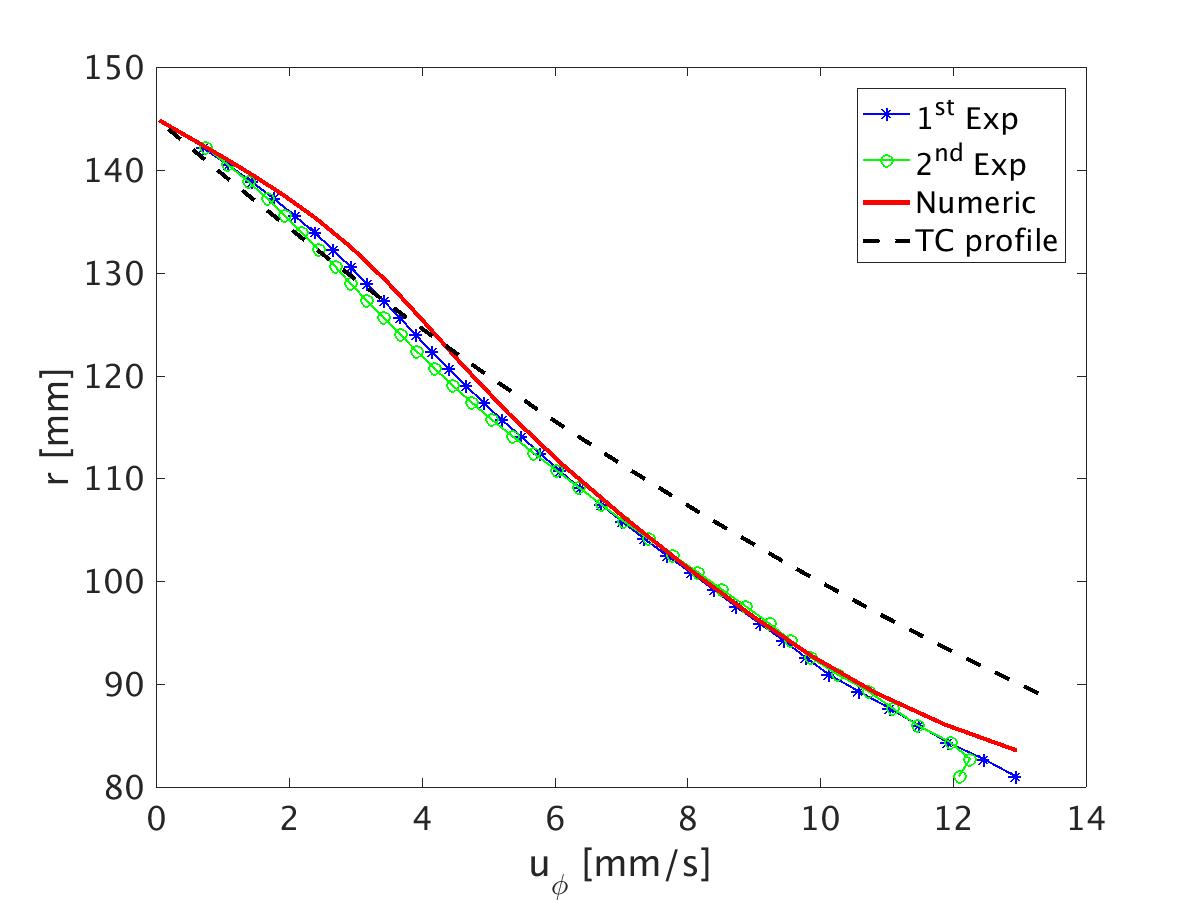}
	\caption{Time average azimuthal velocity profiles ($\overline{u_\phi}$). The lines with square and circle markers represent two different experimental data performed with $Re=400$, $\mu=0.35$ and temperature between top and bottom lids $\Delta T \approx 4K$, leading to $Rn\approx250$ and $Fr\approx1.5$. The full red curve was obtained from numerical simulation, and the black dashed line is the non-stratified TC profile.  \label{fig:U_mean}} 
\end{figure}

Figure \ref{fig:Uphi_Hovmoller_comparison} shows a comparison between numerical and experimental space time diagrams using a 12 minutes time-slice, for Reynolds number $Re=400$ and $\mu = 0.35$ at mid-height axial position ($H/2$). The initial temperature difference imposed between top and bottom lids is $\Delta T \approx 4K$, leading to $\partial T/\partial z \approx 5.7K/m$, $Rn\approx250$ and $Fr\approx1.5$. The reference frame is co-rotating with the outer cylinder for a direct comparison of the results, since the \textit{PIV} experimental data have been obtained in this frame of reference. 
Figures \ref{fig:Uphi_Hovmoller_comparison}.a and \ref{fig:Uphi_Hovmoller_comparison}.b demonstrate the good agreement between numerical simulations and \textit{PIV} data, presenting SRI oscillations with a period of $\approx$~90~seconds in the co-rotating frame of reference. This time scale is relevant to be noticed since it is associated with the SRI most energetic frequency. When compared to the SRI lower frequency oscillations that will be later presented, the SRI frequency is considered to be high.
In figure \ref{fig:U_mean}, the time mean azimuthal velocity profiles $\overline{u_\phi}$ from two different experiments are compared with the corresponding numerical simulations (the bar on top of a variable is used to indicate time average).

The TC profile (\ref{eq:TC}) is represented by a dashed black line. It should be noted that for all SRI profiles observed, the SRI flow is slow near the inner cylinder ($r\approx80mm$) when compared to the non-stratified TC case, and slightly faster near the outer cylinder ($r\approx143mm$), showing how the stratification and the instability affect the mean flow angular momentum transport.

\begin{figure}[tp] 
	\begin{center}
		\begin{minipage}[t]{0.45\textwidth}
			\centering
			\includegraphics[width=1\linewidth]{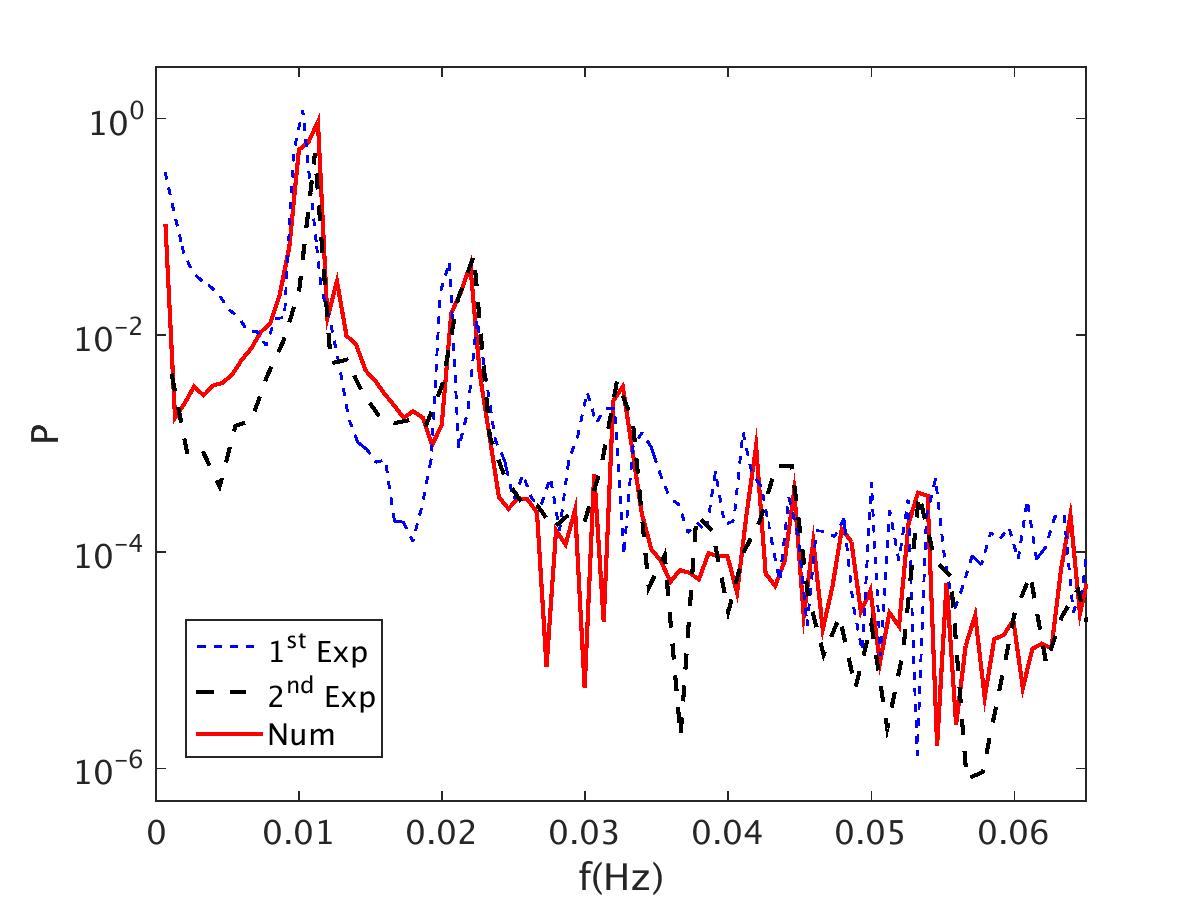}
			\caption*{(a) $Re=400$.}
		\end{minipage}
		\hspace{0.8cm}
		\begin{minipage}[t]{0.45\textwidth}
			\centering
			\includegraphics[width=1\linewidth]{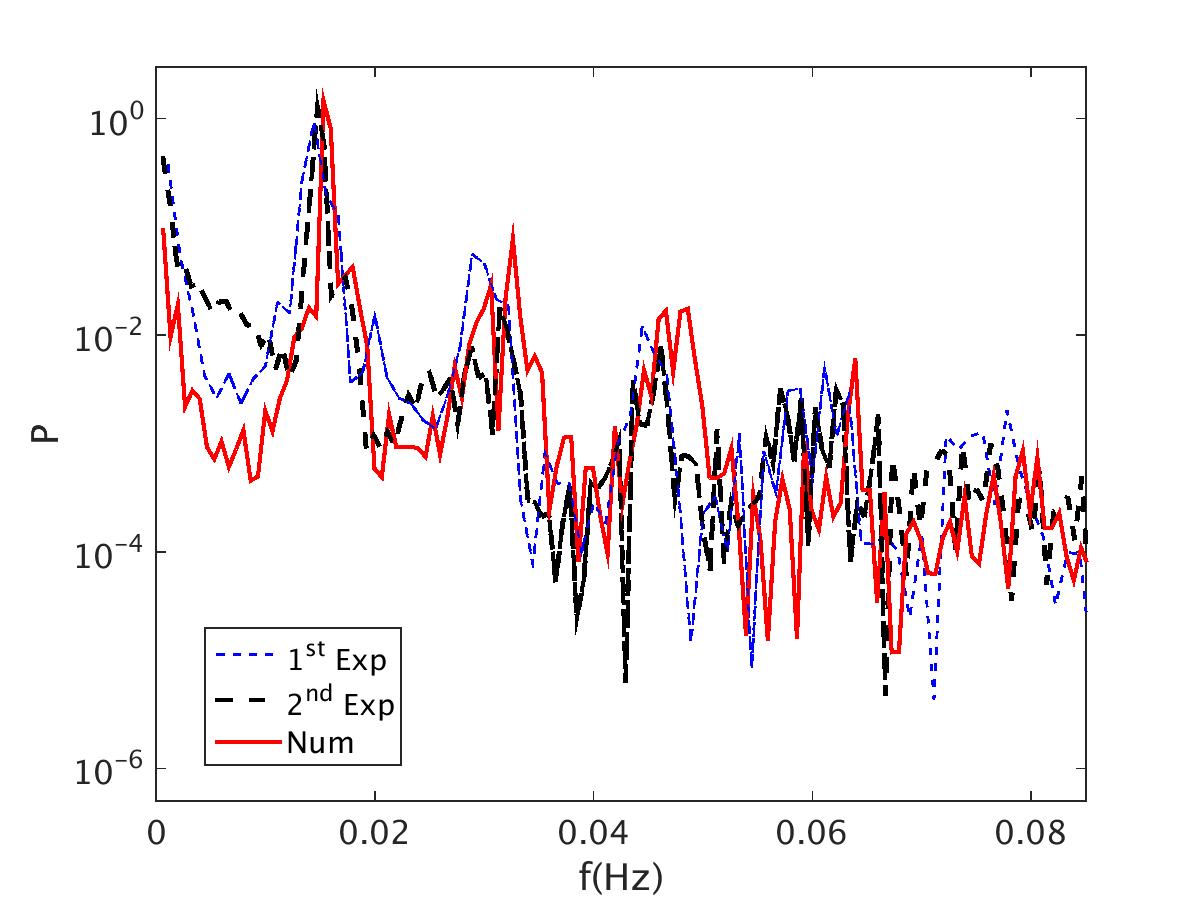}
			\caption*{(b) $Re=600$.} 
		\end{minipage}
		\caption{Comparison of two different experiments and the numerical simulation spectra with amplitude (P) axis shown in logarithmic scale. $\mu=0.35$ and initial temperature between top and bottom lids of $\Delta T = 4K$, leading to $\partial T/\partial z \approx 5.7$K/m, $Rn\approx250$ and $Fr = 1.5$. \label{fig:FFTs_in_logScale} } 
	\end{center}
\end{figure}

Figures \ref{fig:FFTs_in_logScale}.a and \ref{fig:FFTs_in_logScale}.b show a comparison of $u_\phi$ spectra between the two experiments (performed at different days with the same parameters) and the numerical simulation, respectively for $Re=400$ and $Re=600$. The dominant SRI frequencies are $f_{SRI}(Re=400)\approx0.011Hz$ and $f_{SRI}(Re=600)\approx0.015Hz$. The amplitude of the spectra is computed as the square of the Fourier transformed azimuthal velocity ($P =  |{FFT(u_\phi)}|^2 $) at a fixed azimuthal and radial position $\phi=0,r=r_{in}+d/2$, and height $z=H/2$, and it is shown in figure \ref{fig:FFTs_in_logScale}. For both $Re$, the SRI peak corresponds to the mode $m=1$ azimuthal wave number.
The spectra in fig. \ref{fig:FFTs_in_logScale} show good agreement between experimental and numerical data, although the experimental results have the tendency of showing slightly smaller frequencies than the ones obtained numerically, possibly due to the small errors on controlling the experimental inner and outer cylinders velocities. This slightly smaller $\Omega$ imposed in the experiments can be also
noticed in the slightly lower values in the experimental mean velocity profiles of figure \ref{fig:U_mean} when compared to the numerical data. 
Besides showing the good agreement between numerical and experimental data, the log-scale FFTs presented in figure \ref{fig:FFTs_in_logScale} also reveal harmonics of the SRI most energetic frequency, showing that non-linearities are starting to set in for the chosen parameters.
Each mode is connected to the frequencies via
\begin{equation} \label{eq:frame_change}
f_{lab} = f_{rot} + m \Omega_{out}
\end{equation}
where $f_{lab}$ is the frequency in the laboratory frame of reference, $f_{rot}$ is the frequency in an outer cylinder co-rotating frame, $\Omega_{out}$ is the angular velocity of the outer cylinder, $r$ is the radial position at which the time series have been measured. To find out which mode $m$ corresponds to each peak of $f_{lab}$ from the experiment, we can use (\ref{eq:frame_change}) and $f_{rot}$ from the numerical spectrum, since for a chosen m, only one peak of $f_{lab}$ fulfils (\ref{eq:frame_change}), while all the other peaks ($\neq m$) remain uncorrelated.



\subsection{Numerical and experimental observations of the SRI linear stability marginal curves}\label{subsection:comparison_with_linearStability}

\begin{figure}[tp] 
	\begin{center}
		\begin{minipage}[t]{0.45\linewidth}
			\centering
			\includegraphics[width=1\linewidth]{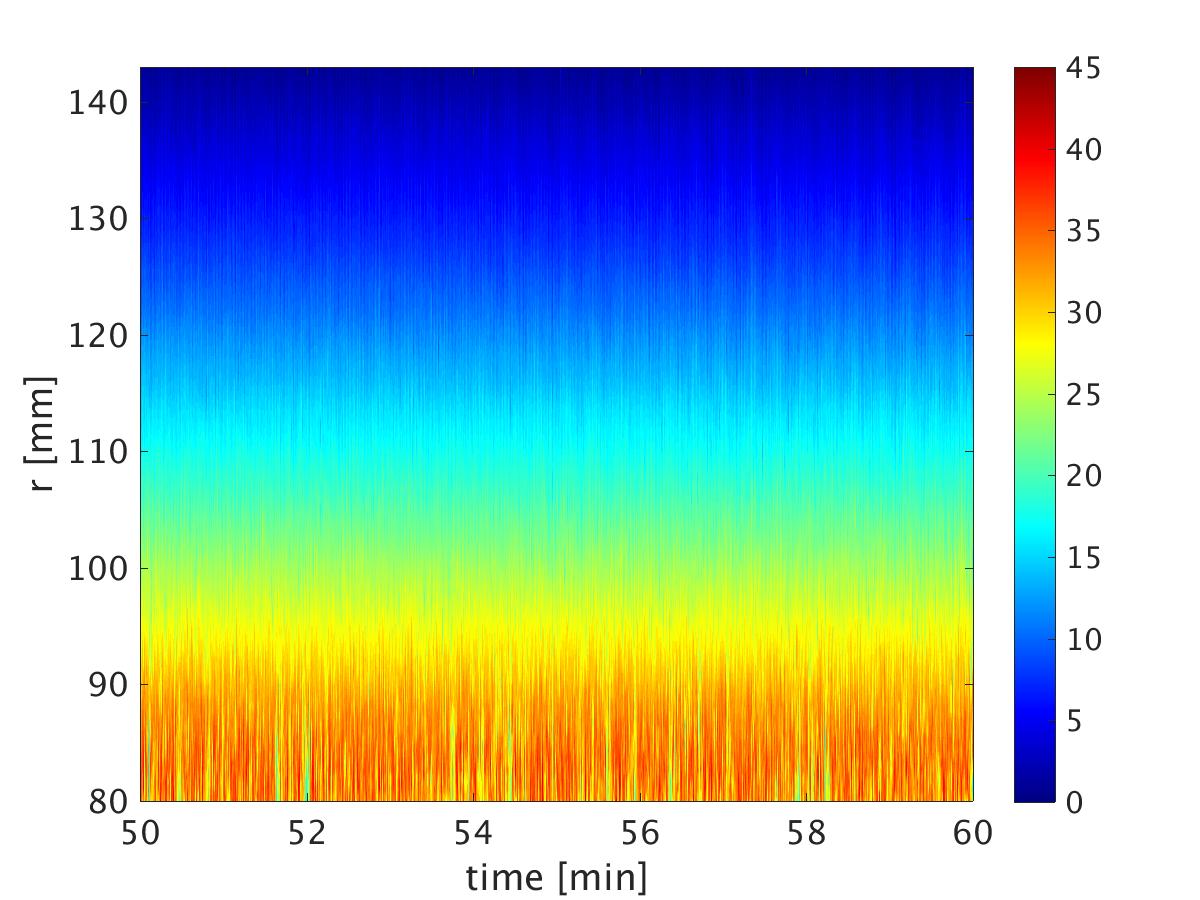}
			\caption*{(a) Experiment} 
			\vspace{4ex}
		\end{minipage}
		\hspace{.1in}
		\begin{minipage}[t]{0.45\linewidth}
			\centering
			\includegraphics[width=1\linewidth]{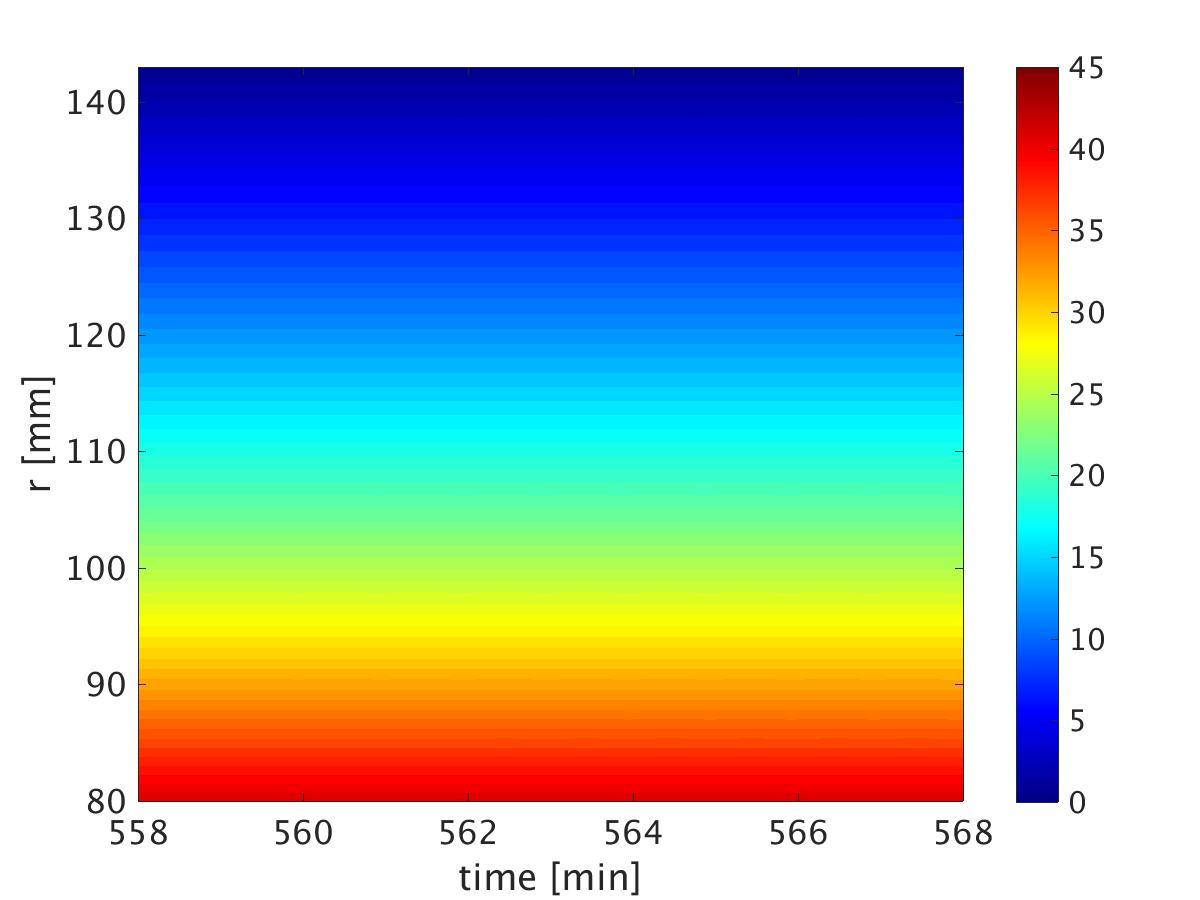}
			\caption*{(b) Numerical Simulation} 
			\vspace{4ex}
		\end{minipage} 
		\caption{$u_\phi$ space time diagrams (Hovm\"oller) showing the SRI stability with $Re=1000$, $\mu=0.35$ and a temperature difference from top to bottom of $\Delta T \approx 4K$.} \label{fig:Uphi_Hov_Re1000} 
	\end{center}
\end{figure}

For Reynolds numbers $Re<1000$, the non-linear numerical model gives the same critical $Re$ as found by \cite{rudiger2017stratorotational} (see Fig. \ref{fig:linear_stability_curve}).
Stable SRI flows -- when the stratification exists, but no SRI oscillations in the space-time diagram are observed -- have the same azimuthal time average velocity profile as a classic TC flow with no stratification. This can be observed in figure \ref{fig:Uphi_Hov_Re1000} for $Re=1000$ with $\mu=0.35$ and a temperature difference from top to bottom $\Delta T \approx 4K$, leading to $Rn\approx250$ and $Fr\approx1.5$.


The symbols in Fig. \ref{fig:linear_stability_curve} show where we found stable and unstable flows experimentally and numerically for $Rn=250$ and $\mu=0.35$. We found agreement with the linear stability analysis by \cite{rudiger2017stratorotational}.

SRI stability for Reynolds numbers larger than $1000$ may be counter-intuitive if we establish an analogy with other turbulent flows driven by shear, as such flows are turbulent above a critical $Re$. However, since the SRI flow we investigated is above the Rayleigh limit, it can be expected that when rotation effect becomes more significant than stratification, i.e., for increasing $Fr$, the physics is approximately the same as non-stratified TC flows and from this point of view it is quite clear that the flow should become stable when $Fr$ becomes too large.

\begin{figure}[tp] 
	\begin{center}
		\begin{minipage}[t]{0.45\linewidth}
			\centering
			\includegraphics[width=1\linewidth]{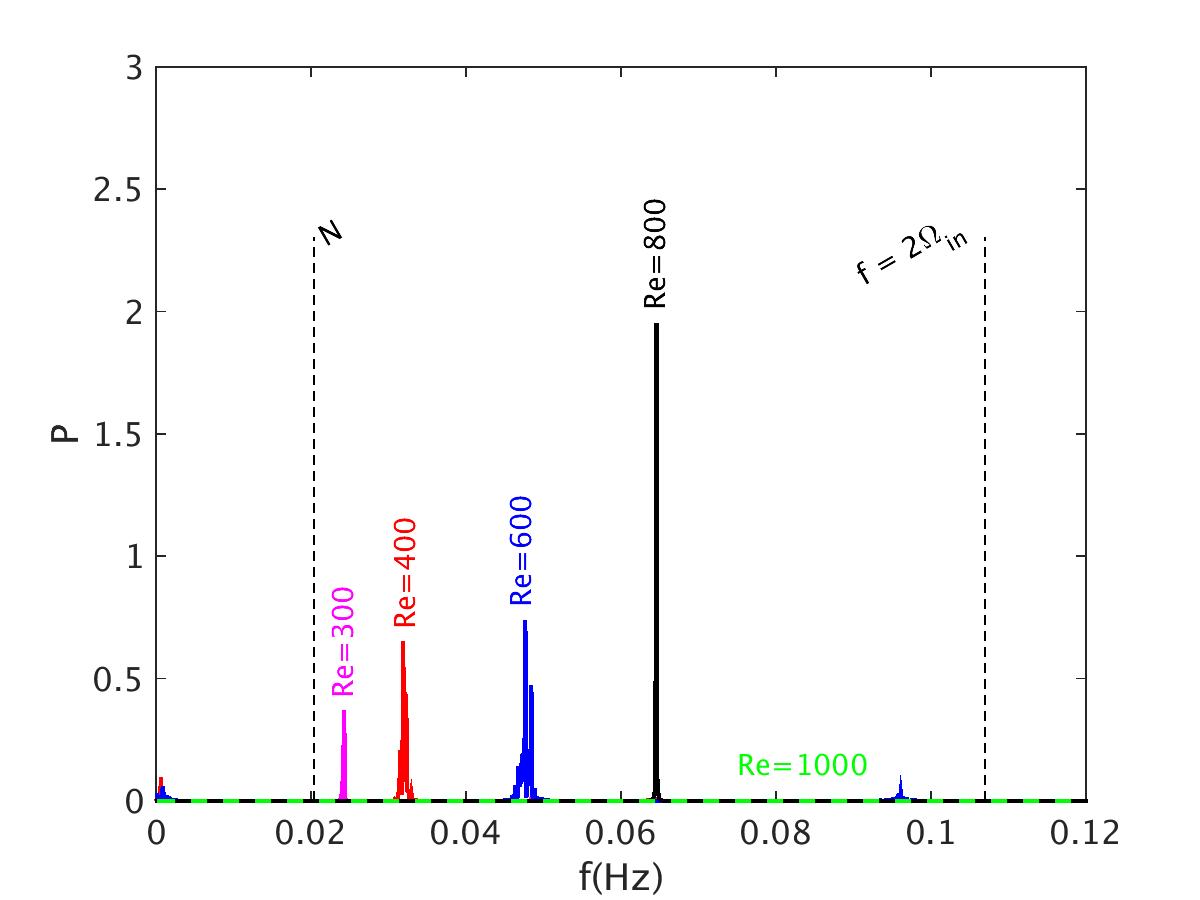}
			\caption*{(a) SRI amplitudes for increasing $Re$. Note that the SRI peak is no longer present in the spectrum for $Re \ge 1000$ (green dashed line at \mbox{$P\approx0$}). The spectra have been taken from a reference frame co-rotating with the outer cylinder.} 
			\vspace{4ex}
		\end{minipage}
		\hspace{.1in}
		\begin{minipage}[t]{0.45\linewidth}
			\centering
			\includegraphics[width=1\linewidth]{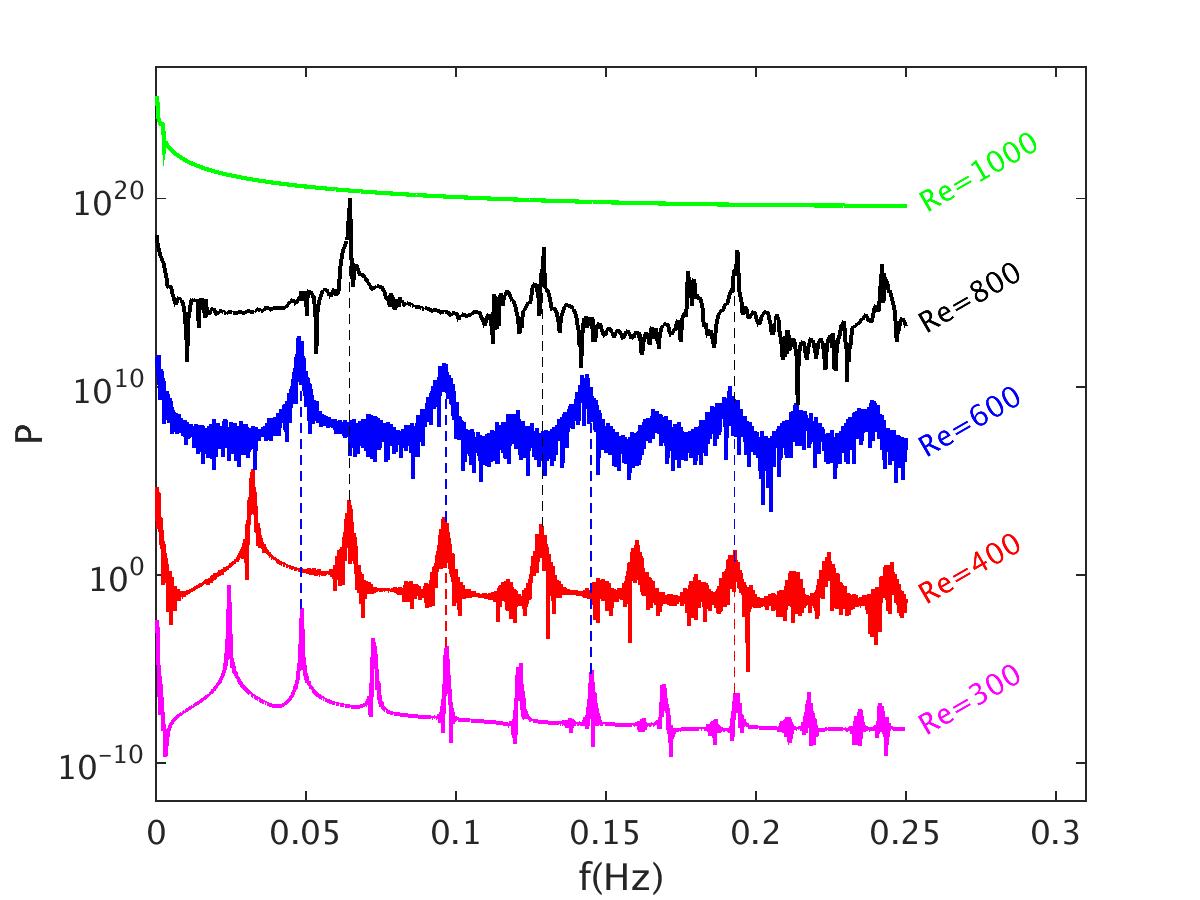}
			\caption*{(b) Spectra for different Reynolds numbers with the amplitude~(P) in logarithmic scale and obtained from $u_\phi$ time series in a laboratory frame of reference. For a better display, the spectra are staggered by multiplying constant exponential functions $c^{te}=e^N$ to displace the spectra vertically in the log scale axis, where $N$ is chosen arbitrarily. Dashed vertical lines are used for highlighting coincident peaks of different spectra.}
			\vspace{4ex}
		\end{minipage} 
		\caption{$u_\phi$ power spectra obtained from numerical simulations at mid-height ($H/2$) and mid-gap ($r_{in}+d/2$) position. The simulations were performed with $\mu=0.35$ and initial temperature difference of $\Delta T = 4K$ ($Rn\approx250$, $Fr\approx1.5$). The horizontal axis shows frequencies in $Hz$, and the vertical axis shows spectra amplitudes $P=|FFT(u_\phi)|^2$. The black dashed vertical line on the left corresponds to the buoyancy frequency for the $Re=400$ case, while the dashed vertical line on the right corresponds to $f=2\Omega_{in}$ for the same case, both corrected by the Doppler shift due to the azimuthal mean flow.   \label{fig:all_FFTs}}
	\end{center}
\end{figure}

Although it seems instructive to see that the SRI peak in the frequency domain increases with the Reynolds number in figure \ref{fig:all_FFTs}.a, where the amplitudes are not presented in logarithmic scale and the SRI peak amplitude is much higher than the amplitudes of its harmonics, this is not true at all different heights in the cavity. In other axial positions, the $Re=600$ spectra can exhibit the most energetic peak, for example.

Figure \ref{fig:all_FFTs}.a highlights how the SRI frequency increases with the Reynolds number until it disappears at $Re=1000$. Note that the SRI frequencies are closer to the buoyancy frequency $N$ than to the inertial frequency $f=2\Omega_{in}$, and they become closer to $f$ when the inner cylinder rotation increases. Furthermore, all peaks in the spectra are inside the interval $N-f$. This is an important remark because internal gravity waves (IGW) can not exist outside the interval $N-f$, where $f=2\Omega_{in}$ (bigger than $N$ in the cases presented). The dispersion relation of IGW with frequencies outside the $N-f$ interval assume complex values \citep{gill1982atmosphere}, therefore, waves cannot exist.   

Figure \ref{fig:all_FFTs}.b shows power spectra for different Reynolds numbers with the vertical axis in logarithmic scale, better showing the harmonics of each SRI spectrum. The spectra in figure \ref{fig:all_FFTs}.b are in a reference frame fixed in the laboratory, since the higher harmonics are better resolved than in the reference frame co-rotating with the outer cylinder as in figure \ref{fig:all_FFTs}.a.

Dashed vertical lines in Figure \ref{fig:all_FFTs}.b are used to highlight harmonics with the same frequencies in spectra of different Reynolds numbers. For example, if
$f_{Re}^j$ is the $j^{th}$ harmonic of an experiment for a certain Reynolds number $Re$ (also equivalent to the m=$n^{th}$~azimuthal mode), then we have
\begin{equation}\label{eq:ReXModes}
    \frac{f^j_{Re}}{f^i_{\frac{j}{i}Re}}=1.
\end{equation}

\noindent For instance, we note that the second harmonic of $Re=300$ coincides with the fist peak of the $Re=600$ spectrum, or that ($f^4_{300}$=$f^3_{400}$ =$f^2_{600}$) and $m_8(Re300)$=$m_6(Re400)$=$m_4(Re600)$=$m_3(Re800)$. 

When the spectra are normalized by the each inner cylinder rotation (therefore, also by the Reynolds numbers), all the frequencies collapse to the same values, i.e., all the spectra become coincident.


\section{Pattern formation}\label{section:low_freq_modulations}

\begin{figure}[tp] 
	\begin{minipage}[t]{0.45\linewidth}
		\centering
		\includegraphics[width=1\linewidth]{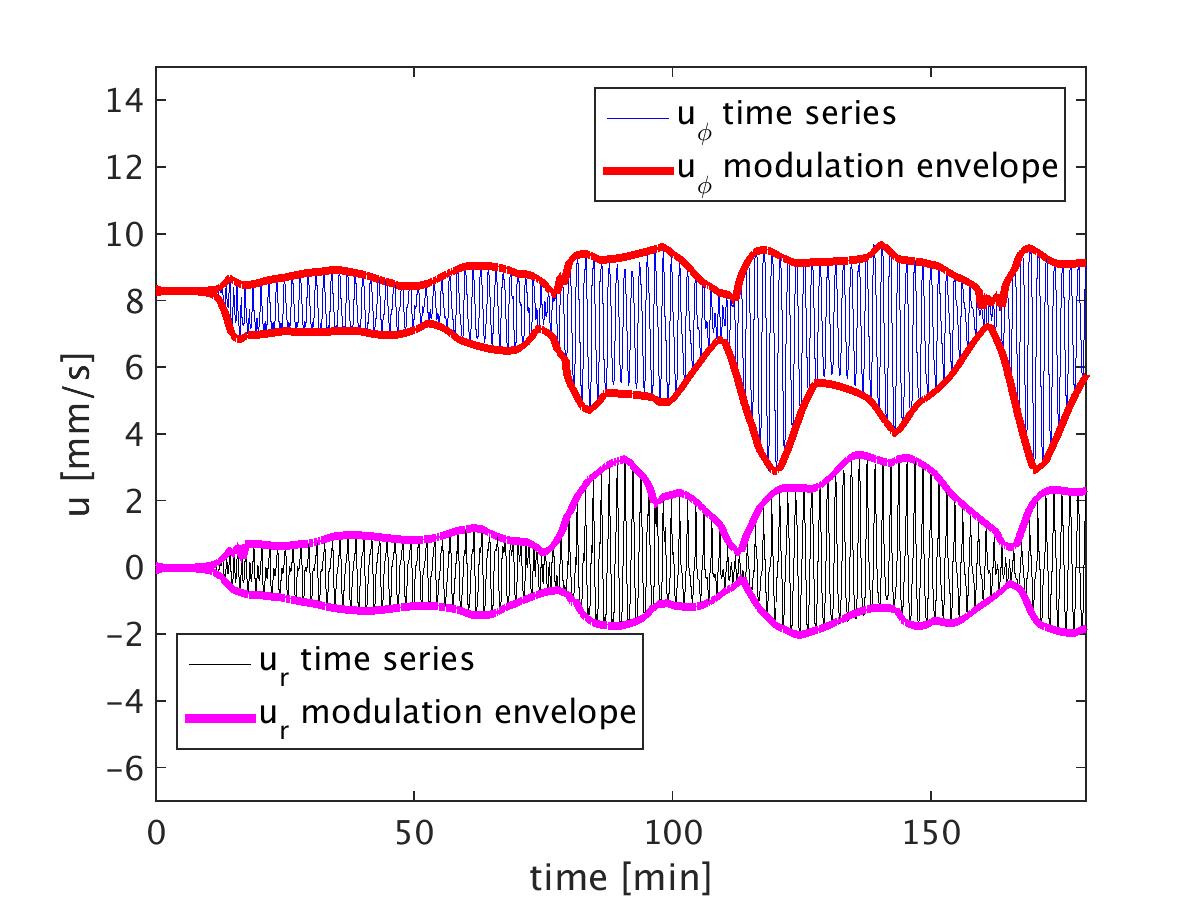}   
		\caption*{(a) SRI time series in the time interval $0<t<3$ hours and their respective amplitude envelopes, highlighting low frequency amplitude modulations.}
		\vspace{4ex}
	\end{minipage}
	\hspace{.1in}
	\begin{minipage}[t]{0.45\linewidth}
		\centering
		\includegraphics[width=1\linewidth]{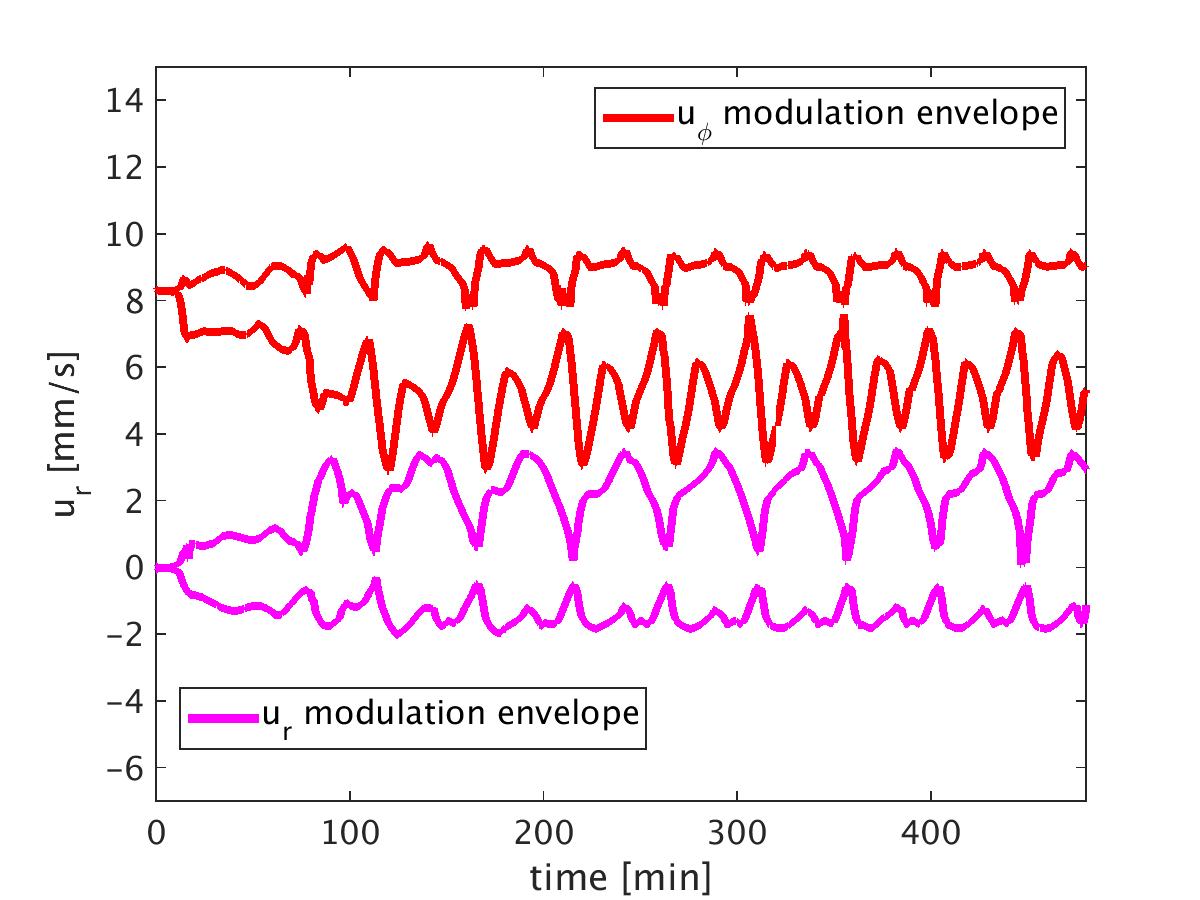}
		\caption*{(b) Velocity amplitude envelopes, highlighting regular low frequency amplitude modulations within the time interval \mbox{$0<t<8$}hours.} 
		\vspace{4ex}
	\end{minipage}
	\caption{Numerical simulation time series for $Re=400$, $\mu=0.35$ and initial $\partial T/\partial z \approx 5.7K/m$ at mid-gap position ($r_{in}+d/2$) and mid height position ($H/2$), and their respective amplitude envelopes highlighting strong low frequency amplitude variations in time. Note that the time intervals are different in figures (a) ($0<t<3$hours) and (b) ($0<t<8$hours), so that the SRI oscillations have been included into figure (a) only.}
	\label{fig:Long_modulation}
\end{figure}

The efficiency of the implemented parallel high performance code described in section \ref{section:Numerical_simulations_desription} reduced long computational times of DNS scalar codes previously employed to investigate the SRI \citep{von2018instabilities,abide20052d,raspo2002spectral,ABIDE2018} from months to hours, which makes it convenient for evaluating low frequency SRI phenomena. When velocity profiles are observed during long time simulations, strong amplitude modulations become evident (figure \ref{fig:Long_modulation}).
These modulations are considered to have low frequency because they are more than 30 times smaller than the SRI frequency. Note that, when the power spectra of the amplitude envelope in figure \ref{fig:Long_modulation} was obtained (not shown here), the low frequency peaks are found outside the interval $N-f$ mentioned previously and therefore, the amplitude modulations cannot be interpreted as low frequency gravity wave modes.

\begin{figure}[tp!] 
	\begin{center}
		\begin{minipage}[t]{0.45\linewidth}
			\centering
			\includegraphics[trim={4cm 0cm 4cm 2cm},clip, width=1\linewidth]{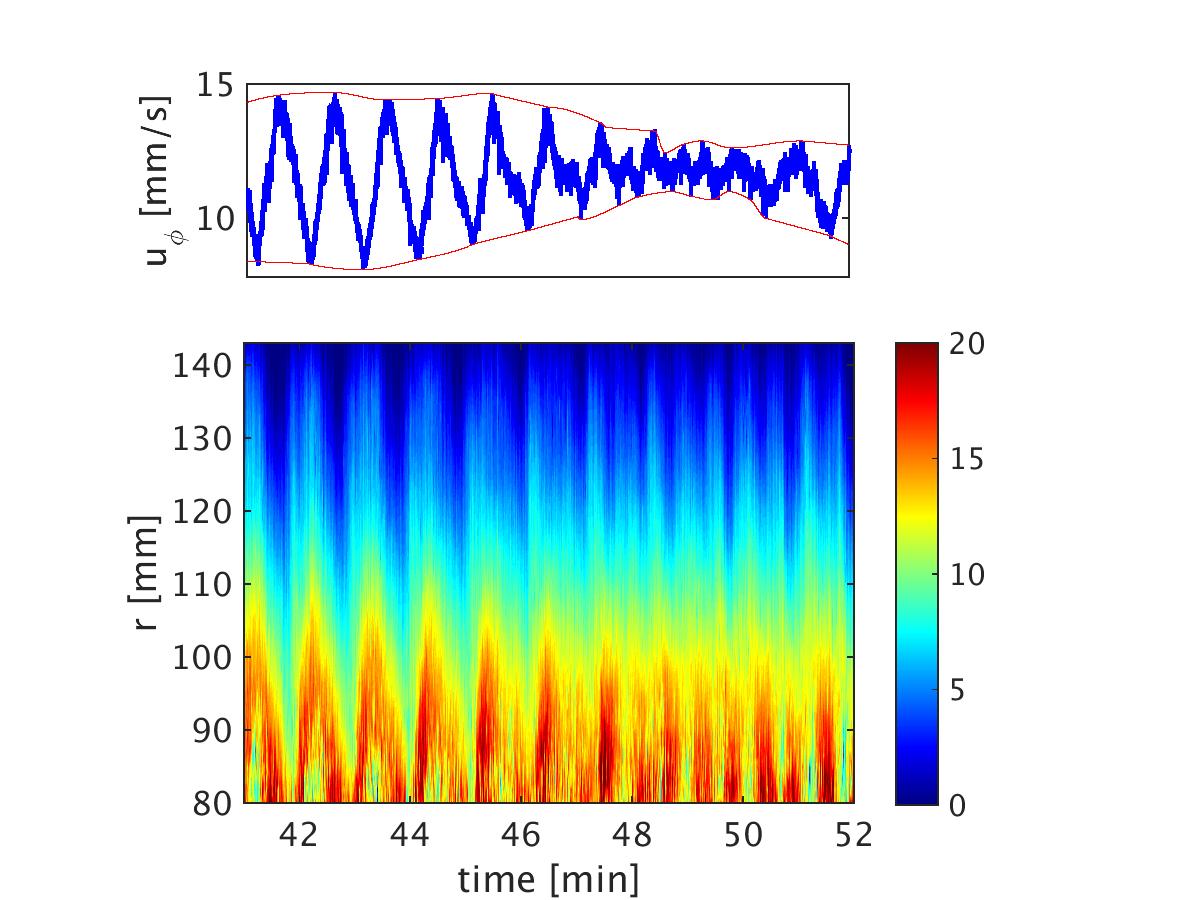}
			\caption*{(a) Experiment} 
			\vspace{4ex}
		\end{minipage}
		\hspace{.1in}
		\begin{minipage}[t]{0.45\linewidth}
			\centering
			\includegraphics[trim={4cm 0cm 4cm 2cm},clip, width=1\linewidth]{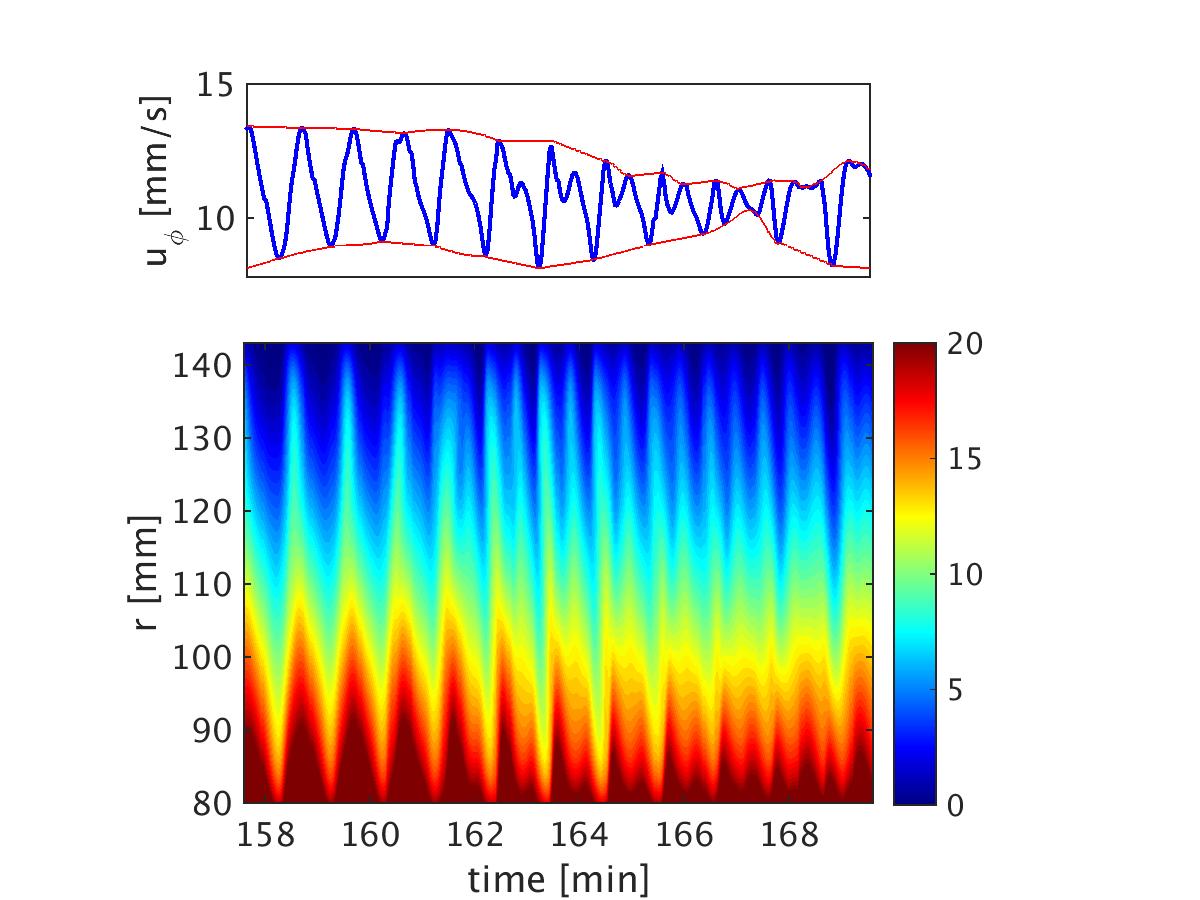}
			\caption*{(b) Numeric} 
			\vspace{4ex}
		\end{minipage} 
		\caption{$u_\phi$ space time diagrams at mid-height axial position ($z=H/2$) during amplitude modulation transition for $Re=600$. The reference frame co-rotates with the outer cylinder, and the velocities represented by the colour scales are given in $mm/s$. The horizontal axis shows time in minutes, and the vertical axis, the radius in mm, where the bottom region is closer to the inner cylinder wall, and the top is closer to the outer one. On top of the space-time diagrams, the velocity profile at mid-gap position ($r_{in}+d/2$) in the space-time diagrams is displayed that highlights its amplitude modulation.} 
		\label{fig:Hov_modulations} 
	\end{center}
\end{figure}

\begin{figure}[!tp] 
	\begin{center}
		\begin{minipage}[t]{0.45\linewidth}
			\centering
			\includegraphics[width=1\linewidth]{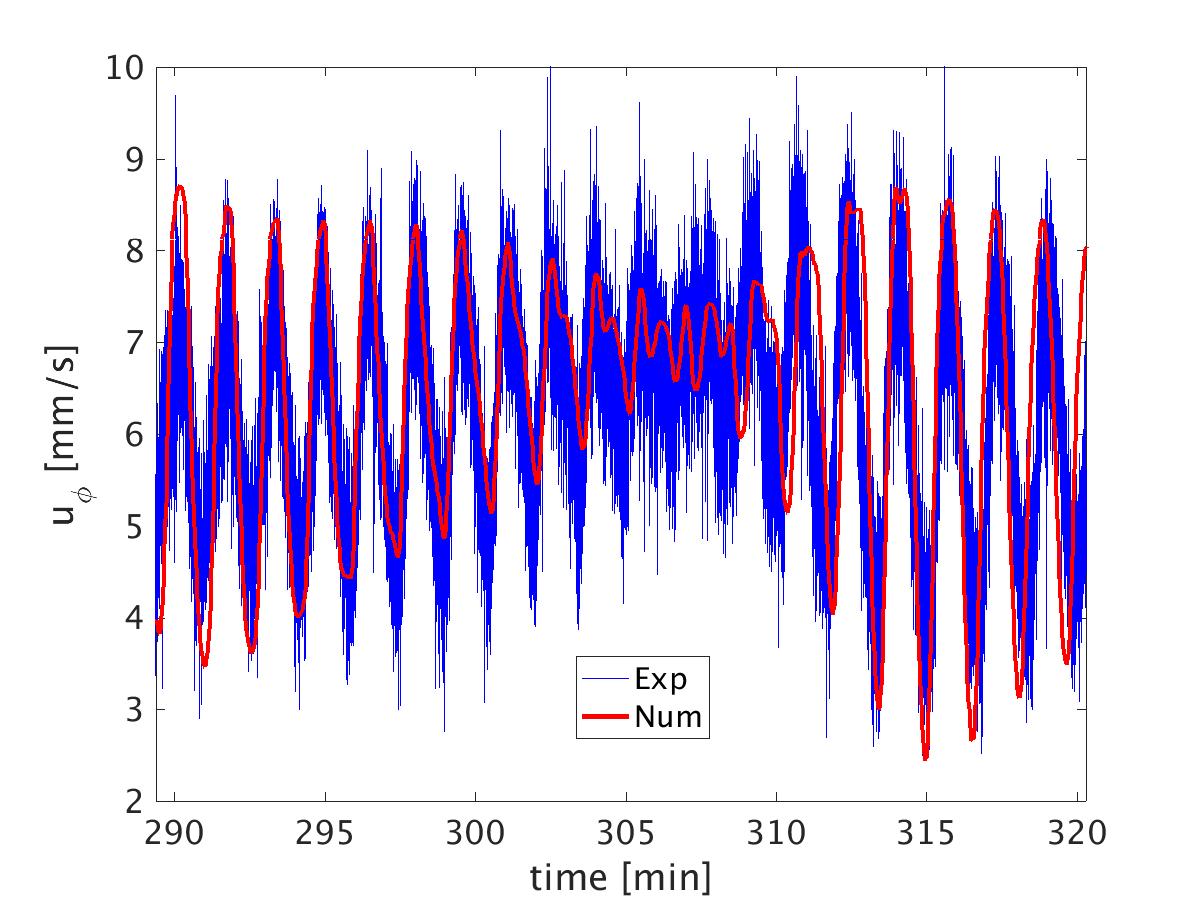}
			\caption*{(a) $Re=400$} 
			\vspace{4ex}
		\end{minipage}
		\hspace{.1in}
		\begin{minipage}[t]{0.45\linewidth}
			\centering
			\includegraphics[width=1\linewidth]{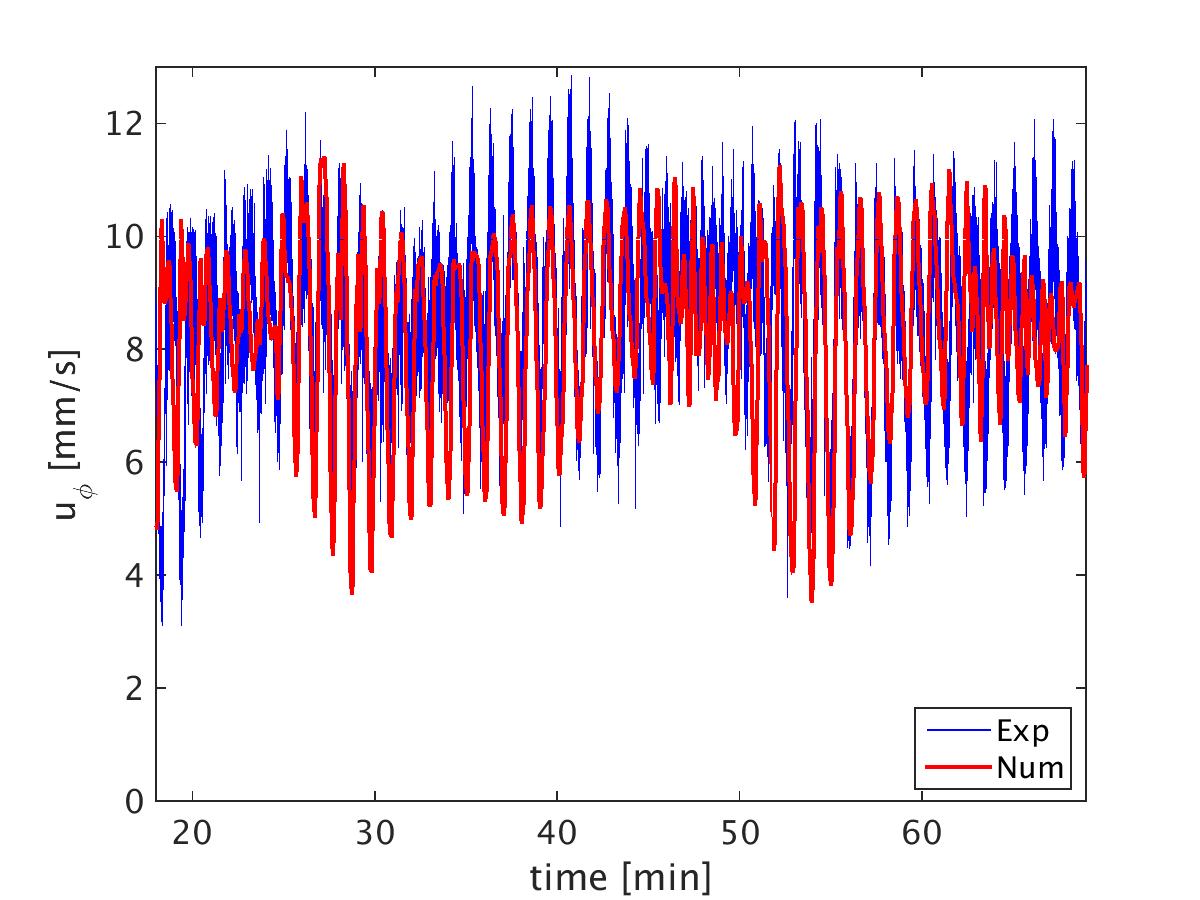}
			\caption*{(b) $Re=600$} 
			\vspace{4ex}
		\end{minipage} 
		\caption{Comparison between experimental ({\color{blue}blue} curve) and numerical simulation ({\color{red}red} curve) $u_\phi$ time series at mid-gap~($r_{in}+d/2$) and mid-height position~($H/2$). Please, note that the time intervals are different in figures (a) and (b).} 
		\label{fig:ExpNum_mudulations} 
	\end{center}
\end{figure}

The modulations shown for numerical simulations in figure \ref{fig:Long_modulation} are also observed experimentally. $u_\phi$ space-time diagrams for $Re=600$ showing amplitude modulation transitions are displayed in figures~\ref{fig:Hov_modulations}~and~\ref{fig:ExpNum_mudulations}. 


These amplitude modulations are observed for all velocity components ($u_\phi$, $u_r$, and $u_z$), for temperature, and hence also in the Brunt-V\"aisäl\"a frequency variations in time. 

It is also possible to see in figure \ref{fig:Long_modulation} that the amplitude variations need approximately 100 minutes to achieve a permanent regime. During the transient regime ($t\lessapprox 100$~minutes), amplitude variations also exist, but they are not regular in time. It is important to highlight that this mentioned transient regime is related to the amplitude modulations, and not to the SRI oscillations. The SRI oscillations show the mode 1 peak in the Fourier space presented in figures \ref{fig:U_mean}.b,  \ref{fig:FFTs_in_logScale}, and \ref{fig:all_FFTs}, and they are already prominent $\approx$ 10 minutes after starting the rotation.

\subsection{Pattern changes associated with the amplitude modulations}\label{subsection:modulations_and_patternChanges}

\begin{figure}[!tp] 
	\begin{center}
		\begin{minipage}[t]{0.45\linewidth}
			\centering
			\includegraphics[width=1\linewidth]{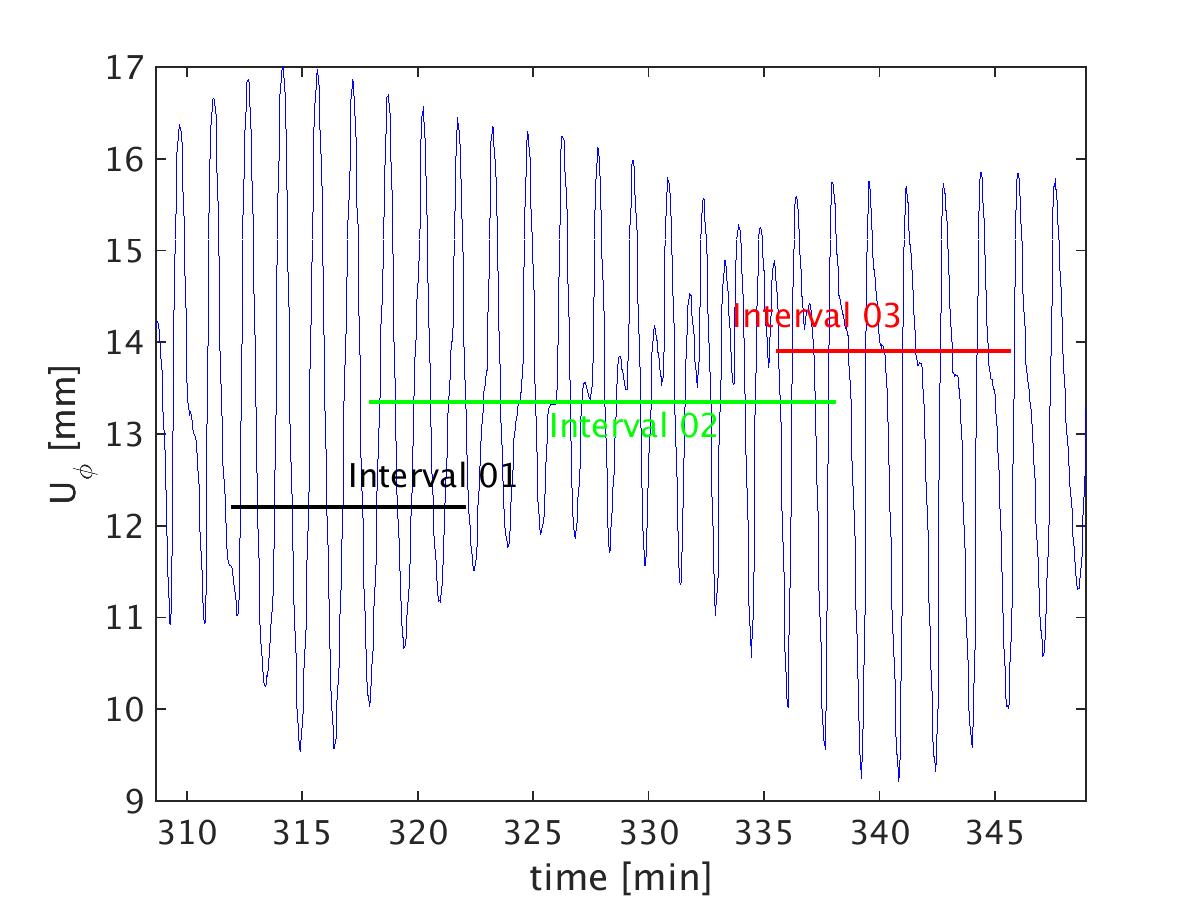}
			\caption*{(a) Time series with horizontal coloured lines indication intervals selected before (\textbf{black}), during ({\color{green} green}), and after ({\color{red} red}) a local minimum amplitude at mid-gap radial position ($r_{in}+d/2$)} 
			\vspace{4ex}
		\end{minipage}
		\hspace{.1in}
		\begin{minipage}[t]{0.45\linewidth}
			\centering
			\includegraphics[width=1\linewidth]{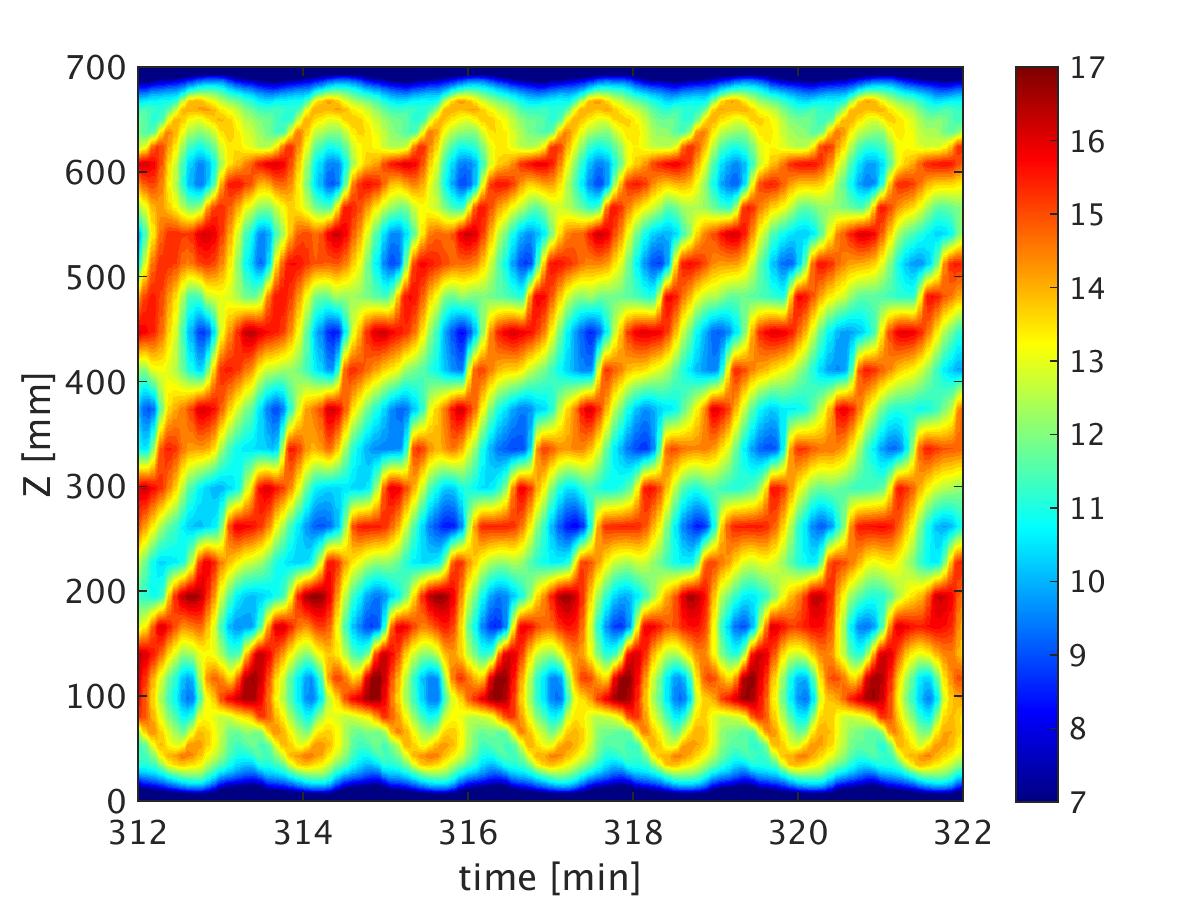}
			\caption*{(b) \textbf{\color{black}Interval 01}, from t = 312 to 322 minutes -- SRI spiral with downward inclination.}
			\vspace{4ex}
		\end{minipage} 
		\begin{minipage}[t]{0.45\linewidth}
			\centering
			\includegraphics[width=1\linewidth]{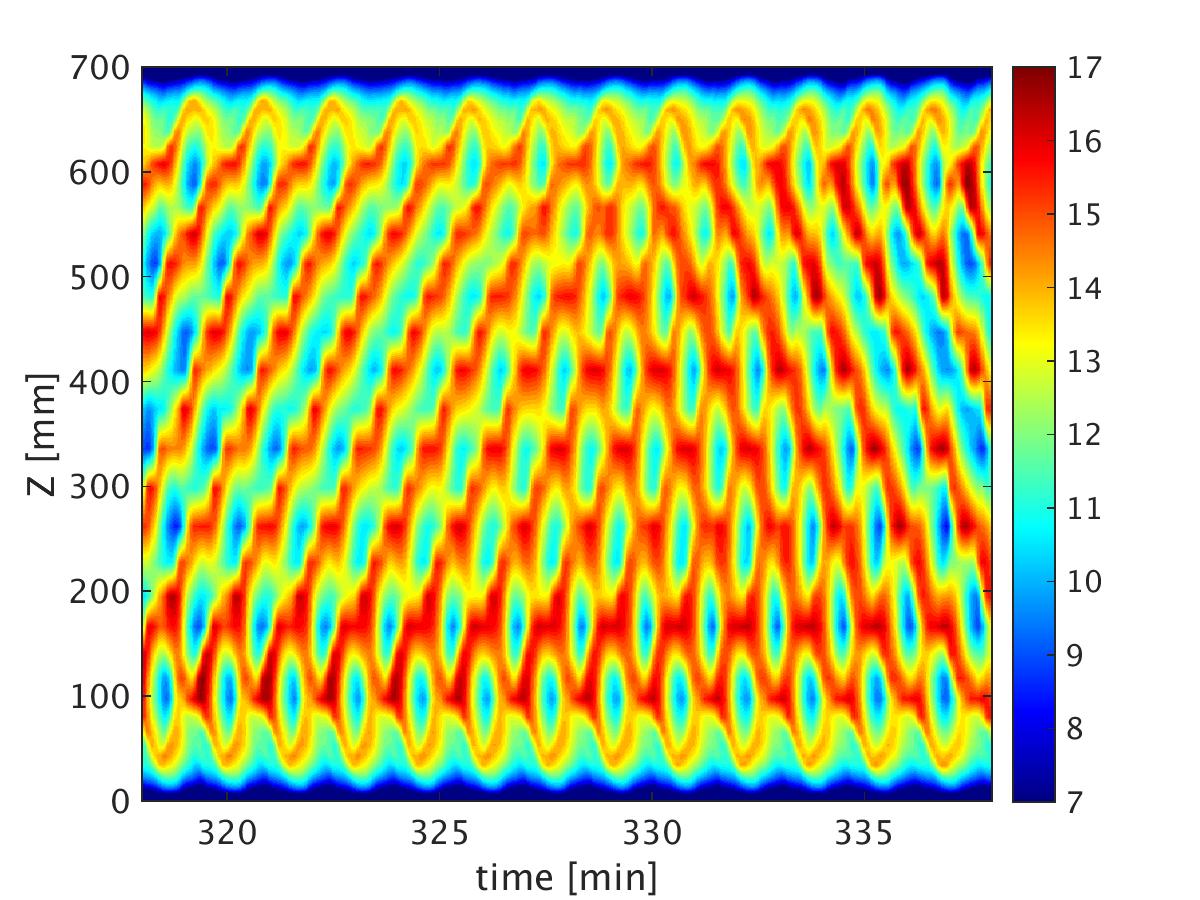}
			\caption*{{\color{green}Interval 02}, from t = 318 to 338 minutes -- transition from a SRI spiral with downward to upward inclination.} 
			\vspace{4ex}
		\end{minipage}
		\hspace{.1in}
		\begin{minipage}[t]{0.45\linewidth}
			\centering
			\includegraphics[width=1\linewidth]{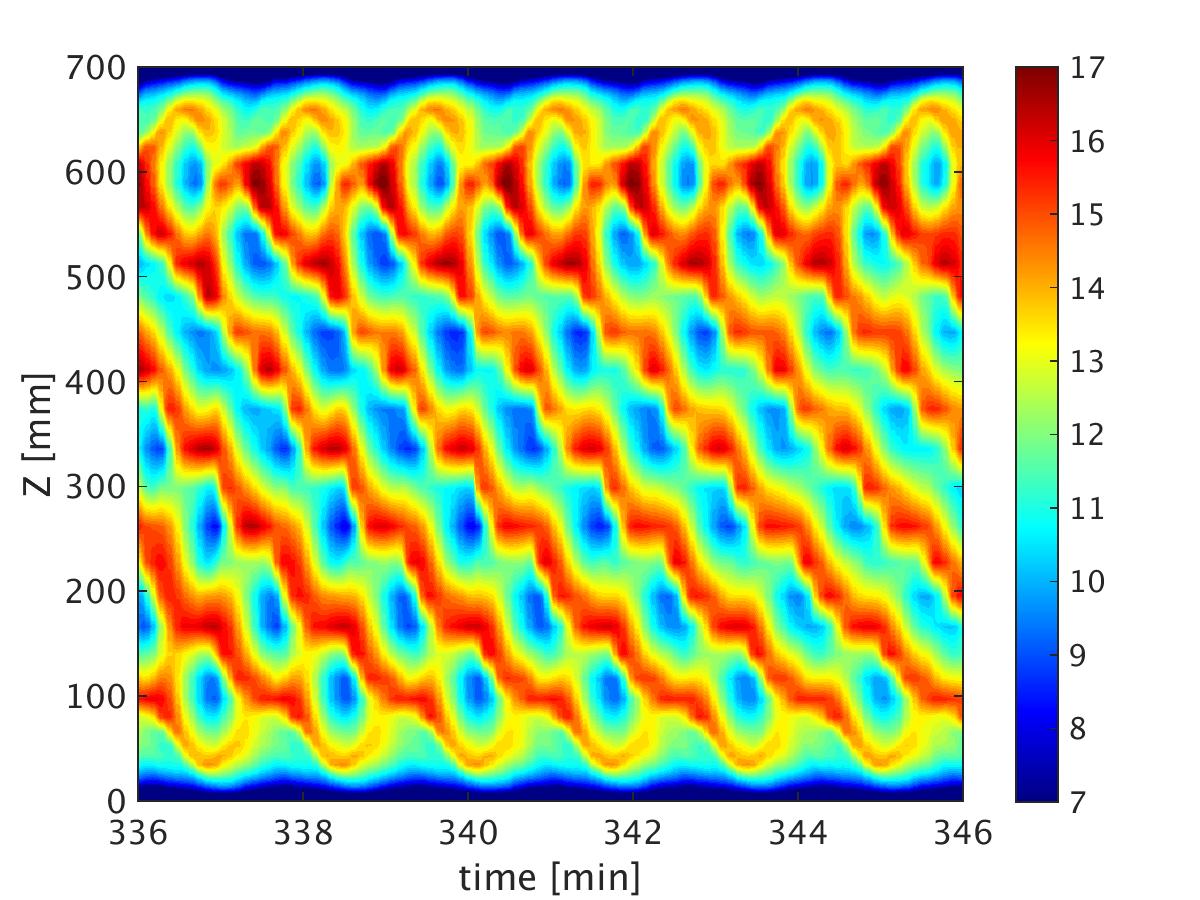}
			\caption*{(d) {\color{red}Interval 03}, from t = 336 to 346 minutes -- SRI spiral with upward inclination.} 
			\vspace{4ex}
		\end{minipage} 
		\caption{$u_\phi$ structures during amplitude modulation.} 
		\label{fig:Pattern_modulation} 
	\end{center}
\end{figure}

The analysis of the SRI flow during the amplitude modulations reveals particular flow patterns that are correlated with the modulations. 
In figure \ref{fig:Pattern_modulation}.a, 3 different time intervals have been selected indicated by colored horizontal lines. In each of these selected intervals, a different flow pattern is observed in the axial-time frame, shown in figures \ref{fig:Pattern_modulation}.b, \ref{fig:Pattern_modulation}.c, and \ref{fig:Pattern_modulation}.d. The patterns represent different SRI spiral inclination and propagation in the axial direction, here named downward inclination (figure \ref{fig:Pattern_modulation}.b) and upward inclination (figure \ref{fig:Pattern_modulation}.d).

During the transition from the upward (downward) to the downward (upward) pattern, both spirals are activated and superposed. This leads to the chessboard type structure pattern in Figure \ref{fig:Pattern_modulation}.c. The transition region is characterized by small SRI amplitudes. 

Observations of the 3 dimensional structures obtained from numerical simulations show that not only the inclination changes during the amplitude modulation, but also does the spiral propagation in the axial direction. The downward spiral inclination (fig. \ref{fig:Pattern_modulation}.b) travels from the top to the bottom lid in the axial direction, while the upward spiral (fig. \ref{fig:Pattern_modulation}.d) travels in the opposite axial direction. 

In the transition region, the SRI spirals do not travel in the axial direction and the spiral angular velocity becomes smaller. It is possible to observe the spiral rotation accelerating in the region where the amplitude grows, and decelerating when the amplitude decreases. Note that the pattern changes also happen in the transition region of non-regular amplitude modulations (before 100 minutes in figure \ref{fig:Long_modulation}).

Although we show the modulations just for $u_\phi$, they can also be observed for $u_r$, $u_z$, temperature, and $N$, but the modulations do not necessarily occur at the same time interval for all these variables. For the case shown in figure \ref{fig:Hov_modulations}, there is approximately 5 minutes of delay between $u_\phi$ amplitude minima and those of $u_r$ or $u_z$ ($u_r$ and $u_z$ pattern transitions occur at the same moment). The reason why the minimum amplitude does not happen at the same moment for the different velocity components was not yet understood, but it indicates that theoretical models to describe these modulations should take into account such amplitude phase shift. The SRI oscillations, otherwise, do not show any phase shift between $u_\phi$, $u_r$ and $u_z$. 

When the SRI amplitude grows, it influences the flow circulation in the $r-z$~plane. This connects the modulations to the study of \cite{hoffmann2009nonlinear} for non-stratified TC flows with small Reynolds
numbers forced additionally by an axial throughflow (with $Re_{axial} = 2$). The TC vortices with the forced axial flow exhibit shape and inclination similar to the ones observed for the SRI, and in general, travels along the axial axis in the same direction as the axial flow imposed, i.e., with the flow in the upward axial direction, the spirals exhibit an upward inclination.
The TC vortices propagation described by \cite{hoffmann2009nonlinear} arises from non-linear defects also observed in our numerical simulations. An example is shown in figure \ref{fig:tecplots}.b, where the defect on the radial velocity profile can be observed near the top lid. 

In the study of \cite{hoffmann2009nonlinear}, a change in the spiral inclination propagating upward occurs when the external axial flow is removed. At this moment, there is a spontaneous break of symmetry leading to a Hopf bifurcation, that implies a change to a downward inclined spiral propagating from the top to the bottom lid, similar to the changes in the SRI spiral we observed. Note that the transition from stable to unstable SRI is also a Hopf-bifurcation \citep{Dubrulle2004}.
One of the differences between the non-stratified TC flows observed by \cite{hoffmann2009nonlinear} and the SRI is that, for the SRI, no external flow in the axial direction is required for the pattern changes to occur, and the SRI spirals are naturally inclined due to the stratification. 
The low frequency pattern changes in the SRI spirals could therefore be interpreted as an oscillation of the system between two slightly unstable fixed points, one fixed point standing for the upward, the other for the downward spiral. 
Furthermore, the stratification values were observed to be important for the secondary instability and the amplitude modulations. By reducing the stratification to half of its value, i.e., when the temperature differences imposed between top and bottom lids were of $\Delta T =2K$ instead of $\Delta T =4K$, keeping all the other parameters constant, the modulation vanishes, although the SRI oscillations still exist, showing that the SRI does not necessarily lead to amplitude modulations.
For an increased initial temperature difference $\Delta T = 8K$, the modulations are again observed. Values of $\Delta T \geq 10$ can no longer be reached with our experimental setup. Moreover, simulations comparing the Boussinesq to low mach number approximation using water as the fluid between the cavities showed that, in this case, the Boussinesq approximation would be no longer valid if $\Delta T$ is larger than $10K$. Therefore, also to guarantee that the Boussinesq approximation is still valid, values of $\Delta T > 8K$ were not investigated. 
Note that the spontaneous break of symmetries in the SRI spiral pattern happens irregularly during the transient phase (time~$<$~100~minutes), and regularly after the modulation patterns are well established.  

We compared the average time that one point in the spiral takes to travel from the bottom to the top of the cavity (or vice-versa, in case of downward pattern) and observed that the axial period of the spiral is at least 10 times smaller than the period of the amplitude modulation (of $\approx 50$~minutes), and no clear correlation was noticed between these two periods. Therefore, since no relation was deduced, the possibility of wave reflecting on the lids to generate the amplitude modulations was discarded. Furthermore, simulations with periodic boundary condition at the bottom and top lids also showed amplitude modulations associated to pattern changes, showing that the presence of lids is not a condition for the modulations and the pattern transitions to occur.

Finally, we note that the same pattern changes associated with the amplitude modulations described here for $Re= 400$ and $\mu = 0.35$ and shown in figure \ref{fig:Pattern_modulation} were also observed for other values of $\mu$ (e.g., $\mu = 0.3752$ -- keplerian line) and for different values of SRI unstable Reynolds numbers.


\section{Conclusions}\label{section:conclusions}

In this paper, we used PIV experimental data and results from a high-performance computing Direct Numerical Simulation (DNS) code, solving the Boussinesq equations, to investigate the Strato-Rotational Instability (SRI). This instability occurs in a fluid between two concentric cylinders rotating at different angular velocities, 
and with a stable density stratification due to a temperature gradient in the axial direction. 

The comparison between numerical simulations and experimental azimuthal velocities $u_\phi$ shows good agreement. The instability manifests itself in the radial-axial plane by oscillations associated to a peak in the spectrum with an azimuthal wavenumber $m=1$. For flows with SRI, the mean azimuthal velocities $\overline{u_\phi}$ become slower than in classic non-stratified stable Taylor-Couette flow in regions near the inner cylinder, and slightly faster near the outer cylinder implying a mean outward momentum flux. 

Both numerical and experimental observations were used to confirm linear-stability curves. It should be noted that increasing moderate Reynolds numbers make the flow return to a stable regime as described by \cite{Ibanez}, \cite{rudiger2017stratorotational}, and Seelig et al. (2018).
The observations of the SRI power spectra indicate that weak harmonic interactions occur between SRI modes, even when the transition to unstable regimes can be well described by a linear stability analysis.

The new high-performance computing numerical code allows for the first time to observe the flow for long time periods which was not possible with a comparatively much slower scalar code, or with short time experiments. This is, in particular, due to the large physical time involved during the establishment of the solutions in SRI configurations, clearly showing the suitability of the present numerical tool to treat such problems. These longer observations revealed that the SRI velocity profiles present strong low frequency amplitude modulations, also observed in the experimental measurements. These amplitude modulations are related to pattern changes of the m=1 SRI mode. In contrast to the most unstable non-stratified Taylor-Couette modes that have the structure of rolls, SRI modes have a nonzero azimuthal wavenumber and hence show a spiral structure (see Figure \ref{fig:tecplots}). A pattern change means
that the spiral inclination in the radial-axial ($r-z$) cross section changes coupled with a change in the direction of the axial drift speed of the spiral.

We can speculate that the observed direction changes result from defects as previously described in the literature on non-stratified TC flows with small $Re$ and forced by an axial perfusion or a circulation in the meridional plane (see e.g. \cite{hoffmann2009nonlinear}).
However, for SRI spirals in the non-turbulent flow regime, the connection to amplitude modulations have not yet been described. The possibility of describing the modulations using the complex Ginzburg-Landau equation is an interesting route we plan to follow in the future \citep{deissler1985noise, bekki1985formations, landamn1987solutions, bartuccelli1990possibility, lopez2020impact}.

The spirals we observe have some similarities with those found in an experiment by \cite{flor2018onset} and discussed in numerical simulations by \cite{lopez2020impact}. In these studies two spirals are moving upward and downward resulting in a standing pattern with low frequency modulations. The latter seems to come from differences in the axial drift speed of the two waves. In our study the drift speed shows some variations but the mean upward and downward speeds are the same. It should be noted that the authors use a very different geometry as in the standard SRI studies done earlier with a radius ratio of just $\eta=1/15$ and $\Gamma=1$, much smaller than, for instance, our values of $\eta=0.52$ and $\Gamma=10$. Moreover, in contrast to our study they use a smaller Froude number $Fr<1$ and a larger Reynolds number $Re>6000$. \cite{lopez2020impact} attribute centrifugal buoyancy to the instability although the relevant term in the equations is small and, for our experiments and simulations,
centrifugal buoyancy is even one order of magnitude smaller than in \cite{lopez2020impact}. We hence assumed that centrifugal buoyancy does not play a significant role for the experiments and, in fact, it was not included in the numerical model.

Finally we mention that our experimental setup gives us the opportunity of exploring also counter-rotation SRI regimes. Although counter rotating regimes may have no direct application in the accretion disk theory, it can also give relevant information about the physics of the SRI \citep{park2018competition}, therefore, these regimes are planned to be explored in a future project.

\section*{ACKNOWLEDGMENTS}

Gabriel Meletti and Uwe Harlander acknowledge the financial support from the DFG core facility
center 'Physics of rotating fluids', DFG HA 2932/10-1, and from the Graduate School of BTU Cottbus-Senftenberg and of the Aix Marseille University. Further, the support of a French-German cotutelle program by Aix Marseille University and BTU Cottbus-Senftenberg is gratefully acknowledged. 
The authors thank Isabelle Raspo and Anthony Randriamampianina for many helpful discussions and constant support of our work and Torsten Seelig for help with the data analysis. We further thank Ludwig Stapelfeld, Robin St{\"o}bel, and Vilko Ruoff for technical support.

\appendix

\section{Experimental procedure} \label{Appendix:Exp_procedure}

The process of heating and mixing the flow (by fast rotating the inner cylinder) takes between 2 and 5 hours, which makes measurements more time consuming compared to salt stratified experiments.
After a quasi-linear temperature profile is established in the axial direction, the outer cylinder is rotated while the inner cylinder is kept at rest to eliminate perturbations generated during the temperature mixing process. After stopping the outer cylinder rotation, we wait at least 20 minutes before starting each measurement to guarantee that the flow is at rest. No relevant heat losses or changes in the temperature linearity are observed at this moment. This is due to the fact that there is almost no convective fluid motion, and the heat transfer happens mainly by conduction, which is a slow process with the used M5 oil. 

For cooling the bottom end-plate of the experimental setup, a hose is connected to the lid and attached to an external cooler. Cold water leaving the cooler is pumped to the hose, removing heat from the bottom part of the experimental apparatus. 

\begin{figure}[tp] 
	\begin{center}
		\centering
		\includegraphics[width=.5\linewidth]{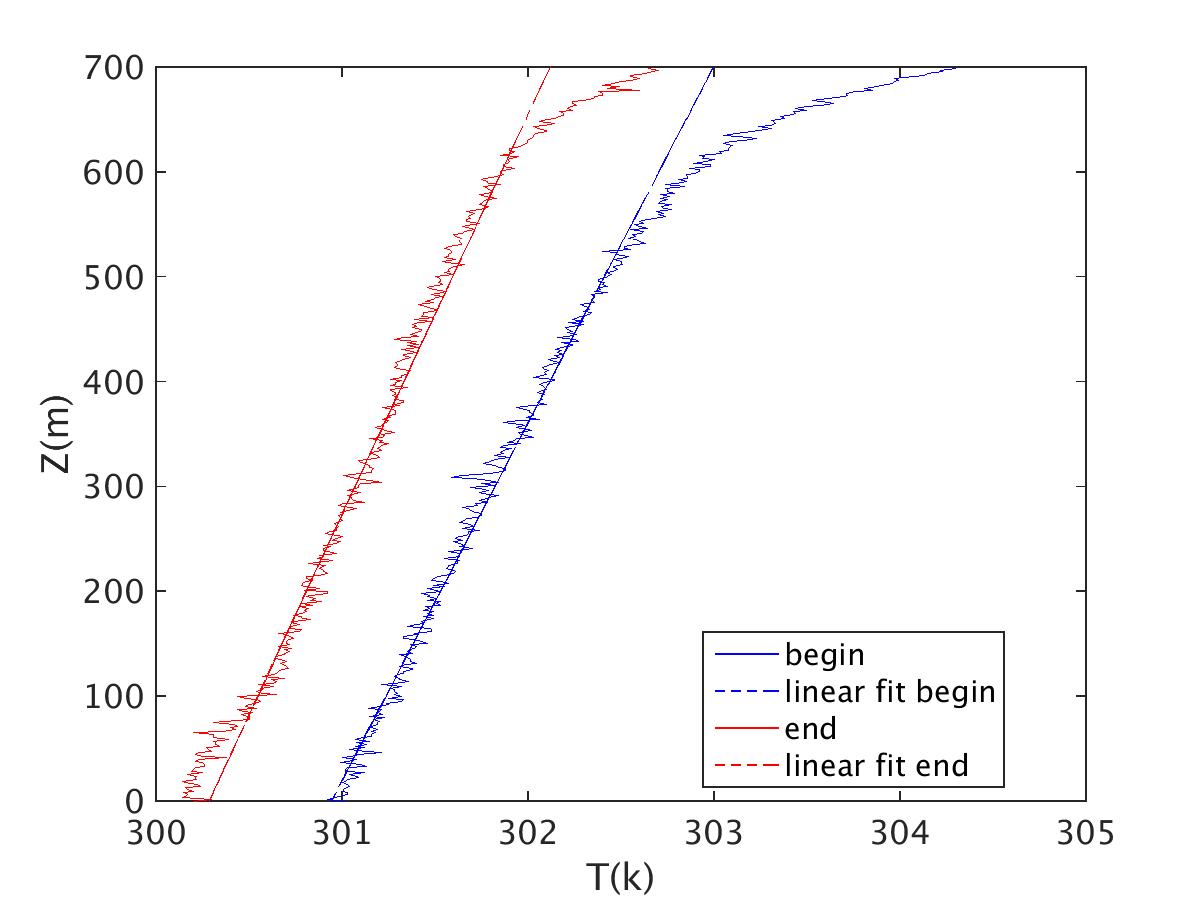}
		\caption{Experimental temperature profiles at the beginning (blue curve on the right) and at the end (red curve on the left) of a measurement. The dashed lines is the linear fit in the center height region where the PIV measurements are performed.} \label{fig:Temperature_Exp} 
	\end{center}
\end{figure}

Since the bottom plate is connected to the outer cylinder, and the hose is attached to the external cooler, that is fixed at the laboratory, the experimental setup cannot be cooled while the outer cylinder is moving.
The Peltier elements for heating can still work while the experiment is running, but this makes the temperature control less precise. Therefore, the fluid might heat up during the experiment, but without affecting much the temperature gradient. Each experiment typically runs between 30~and~70~minutes.
The temperature values along the vertical axis are measured with a PT-100 (Platinum Resistance Temperature Detector) probe, that has a precision $\delta T = 0.1K$. The temperature profile values and its linearity are also verified with an infrared camera IR-TCM 640hr, of $640 px\times480 px$ resolution. The maximum temperature differences between top and bottom lids obtained at the beginning of the experiment are of $\Delta T = 7K$, but in our measurements, it is kept between $3K<\Delta T< 4.5K$. An example of the temperature profiles before and after the measurements is shown in figure \ref{fig:Temperature_Exp}.

After the stable linear temperature gradient in the axial direction is established, the experiment is started by rotating the outer cylinder until its final $\Omega_{out}$ value, avoiding initial perturbations that would occur by starting the experiment rotating the inner cylinder. When the outer cylinder reaches the rotation it will have during the experiment, the inner cylinder rotation is gradually increased from rest until the desired $Re$ and $\mu$ values are achieved.

The camera and the laser sets are mounted on a structure that co-rotates with the outer cylinder, so all experimental results are obtained in a reference frame moving with angular velocity $\Omega_{out}$ with respect to the laboratory frame of reference. The energy for the PIV system is provided by 2 charged power banks.
After finishing the PIV measurements, the temperature profile is once again measured. In a one hour experiment, we observe losses of $\approx 1K$ in the $\Delta T$ initially established due to temperature mixing in the axial direction and to losses into the surrounding environment. In spite of these
temperature changes, the quasi-linearity of the profile is kept, and changes in $N$ values are considered small.  

Although it is difficult to determine accurately the uncertainty associated to each of our measurement steps, it is possible to compute the final PIV error of $\epsilon\approx 2\%$ by comparing non-stratified stable TC azimuthal velocity measurements with analytical solutions. A more detailed description of our error estimation can be found in \cite{seelig2018experimental}.

The Reynolds number values investigated here are between $300<Re<1300$. Representative results presented in section \ref{section:Results} are for values of $Re=400$, $Re=600$ and $Re=1000$, respectively corresponding to inner cylinder angular velocities of $\omega_{in}(Re=400)\approx0.381$~rad/s, $\omega_{in}(Re=600)\approx0.571$~rad/s, and $\Omega_{in}(Re=1000)\approx0.952$~rad/s.

\bibliographystyle{apalike}

\bibliography{mybib_02.bib}

\medskip


\end{document}